\documentclass[twocolumn]{aastex61}

\newcommand{\msun}{M_\odot}
\newcommand{\rsun}{R_\odot}
\defcitealias{RH14}{H14}

\received{\today}
\revised{October 27, 2017}
\accepted{\today}

\submitjournal{ApJ}

\shorttitle{Comprehensive study of ejecta-companion interaction}
\shortauthors{Hirai et al.}

\begin{document}

\title{Comprehensive Study of ejecta-companion interaction for core-collapse supernovae in massive binaries}

\correspondingauthor{Ryosuke Hirai}
\email{ryosuke.hirai@physics.ox.ac.uk}

\author{Ryosuke Hirai}
\affil{Department of Physics, University of Oxford, Keble Rd, Oxford, OX1 3RH, United Kingdom}

\author{Philipp Podsiadlowski}
\affil{Department of Physics, University of Oxford, Keble Rd, Oxford, OX1 3RH, United Kingdom}
\affil{Argelander-Institut fuer Astronomie der Universitat Bonn, Auf dem Huegel 71, 53121 Bonn, Germany}

\author{Shoichi Yamada}
\affil{Faculty of Science and Engineering, Waseda University, 3-4-1, Okubo, Shinjuku, Tokyo 169-8555, Japan}

\begin{abstract}
 We carry out a comprehensive study of supernova ejecta-companion interaction in massive binary systems. We aim to physically understand the kinematics of the interaction and predict observational signatures. To do this we perform simulations over a vast parameter space of binary configurations, varying the masses of the progenitor and companion, structure of the companion, explosion energy, and orbital separation. Our results were not so consistent with classical models by \citet{whe75}, sometimes deviating by an order of magnitude. We construct an alternative simple model which explains the simulated results reasonably well and can be used to estimate impact velocities for arbitrary explosion profiles and companion star structures. We then investigate the long term evolution after the supernova, where the companion can be inflated by the energy injected into the star. We find that the companion can become more than an order of magnitude overluminous straight after the supernova, but quickly fades away after $\sim10$\,years and returns to its original luminosity in about a thermal timescale of the star. Finally, we also discuss the possible surface contamination of heavy elements from the slower ejecta. 

\end{abstract}

\keywords{binaries:close---hydrodynamics---stars:massive---supernovae:general}

\section{Introduction} \label{sec:intro}
Massive stars with masses $\gtrsim8\,\msun$ are known to experience core-collapse (CC) supernovae (SNe) at the end of their lives leaving compact remnants such as neutron stars (NSs) and black holes (BHs). There are many indications from recent surveys that these massive stars form mainly as members of binary or higher order multiple systems \citep[e.g.][]{chi12,san14}, and many of them have close enough orbital separations so that the stellar components interact during their evolution \citep[]{san12,san13,sch14}. Therefore the majority of CCSNe should be occurring in massive close binaries. 

The outcome of the explosion determines the further evolution of the binary. In some cases the binary can be disrupted, so the companion star and the compact remnant would carry on their evolutions as single stars. In other cases, the system should remain bound after the SN, continuing its evolution as a binary. Both cases have been identified by observing stars in supernova remnants (SNRs). For example, a disrupt binary has been detected in the SNR S147 \citep[]{din15} and some other nearby SNRs \citep[]{bou17}. On the other hand, some X-ray binaries have been found inside SNRs \citep[e.g.][]{hen12,sew12}, indicating that the binary has remained bound. There is also an interesting case where a polluted solar-type companion to an X-ray source was found in an SNR \citep[]{gva17}. The subsequent evolution of the surviving binaries will have a rich diversity depending on their masses and binary parameters (orbital separation, eccentricity, etc). If the orbital separation is sufficiently small, the secondary will evolve to fill its Roche lobe and then transfer mass to the remnant of the primary or trigger a common-envelope phase. If the secondary-star mass is large enough, the system may even experience a second SN. Such cases are important for understanding the formation of double NSs, which have attracted wide attention since the detection of gravitational waves from a binary NS merger event \citep[]{abb17}.

After a SN explodes in a binary, the gravitational bond between the two components weakens due to the sudden mass loss. The binary will become unbound if more than half of the total mass is expelled instantaneously. Even if the system survives the explosion (i.e. does not become unbound), the energy and momentum imparted to the companion by the SN ejecta can strip some extra mass or give a final push to unbind them \citep[]{whe75}. Such effects have been dubbed as ``ejecta-companion interaction (ECI),'' and has been intensively studied in the context of type Ia SN \citep[e.g.][]{frx81,liv92,mar00,men07,kas10,liu12,liu13,liu15a,pan13,sha13,mae14,nod16}. In contrast, there are very few studies that have focused on the effect of ECI for the case of CCSNe \citep[]{RH14,liu15b,RH15,rim16}.

\citet[]{RH14} attempted to derive an upper limit to the amount of mass stripped off by ECI. They simulated the effect of the collision of SN ejecta on a $10\,\msun$ red-giant companion, which has a very loosely bound envelope. Approximately $\sim25\,\%$ of the envelope was removed, which is enough to help destroy the binary. However, it is very unlikely that the companion is an evolved star for CCSNe, so their results were not applicable to realistic situations. \citet{liu15b} and \citet{rim16} both carried out ECI simulations on low mass companions ($0.9-3.5\,\msun$), with 3D smooth particle hydrodynamic (SPH) codes. Both studies extensively studied the total unbound mass and final momentum of the remaining star, and their dependences on explosion properties and orbital separation. However, the masses of the companion star models used in their simulations are limited by the numerical cost, and were not able to study ECI with massive companions.

One important conclusion that many of the previous simulations have reached is that the amount of unbound mass and the impact velocity depend strongly on the structure of the companion star. For example, all studies show that values of the total unbound masses or impact velocities follow a power law of the orbital separation. The values of the power differ between the companion models used, and it was not clear what determines it. 

In this paper we will comprehensively investigate the effects of ECI on a massive companion. We carry out two-dimensional hydrodynamical simulations systematically over a wider parameter space of binary parameters than previous studies. We especially aim to understand how the stellar structure actually determines the outcomes of ECI. We also discuss some possible observational anomalies due to ECI. The numerical method and stellar model used will be outlined in Section \ref{sec:numerical}. Results of the simulations will be presented in Section \ref{sec:results} along with results of convergence tests. Based on our results, we will particularly focus on the momentum transfer efficiency from the ejecta to the companion in Section \ref{sec:momentum_transfer} and its influence to the resulting orbit in Section \ref{sec:orbit}. Possible observational signatures of ECI will be discussed in Section \ref{sec:observational}. We will summarize and conclude our claims in Section \ref{sec:conclusion}.

\section{Numerical Method}\label{sec:numerical}
In this section we describe the numerical methods that were used in this study. To simulate the whole process of SN ejecta colliding with the companion star in a binary, we adopt the same two-step strategy that was taken in \citet{RH14}. We first hydrodynamically simulate the explosion of the primary star in spherical symmetry. Then in the second step we simulate the collision of the SN ejecta with the companion star in axisymmetry.
All hydrodynamical simulations were carried out using the code developed in \citet{RH16}. It solves the ideal magnetohydrodynamic equations with the finite volume method, with HLLD-type approximate Riemann solvers for the numerical flux \citep[]{miy05}. Since we do not include magnetic fields in this study, it is equivalent to solving the Euler equations using HLLC-type fluxes. For multidimensional simulations, self-gravity is treated in an original way using the ``hyperbolic self-gravity solver,'' where the gravitational field is evolved by a wave equation (a hyperbolic partial differential equation) instead of solving the Poisson equation (elliptic partial differential equation) at each time step. This dramatically reduces the numerical cost, enabling our wide systematic study. Details of the methodology and stellar models used in each step are given below.

\subsection{Step 1: Explosion of the Primary Star}\label{sub:method1}
The aim of our first step is to obtain the structure of the SN ejecta. One of the key difficulties for modelling CCSN ejecta is that there is a large diversity compared to type Ia SNe. Ejecta masses may spread from fractions of a solar mass to tens of solar masses. Explosion energies can also range from $10^{50}$\,erg for weak explosions to $10^{52}$\,erg for hypernova-type explosions. Even for the same mass and explosion energies, the momentum distribution in the ejecta may differ depending on the structure of the progenitor star. It is clear that all these aspects are closely correlated with each other. However, the correlation between SN progenitor features and explosion energy are poorly understood despite numerous efforts \citep[e.g.][]{nak15,yam16,suk16}. Here we simply create two representative progenitor models and treat the explosion energy as a free parameter.

\begin{figure}
 \centering
 \includegraphics[]{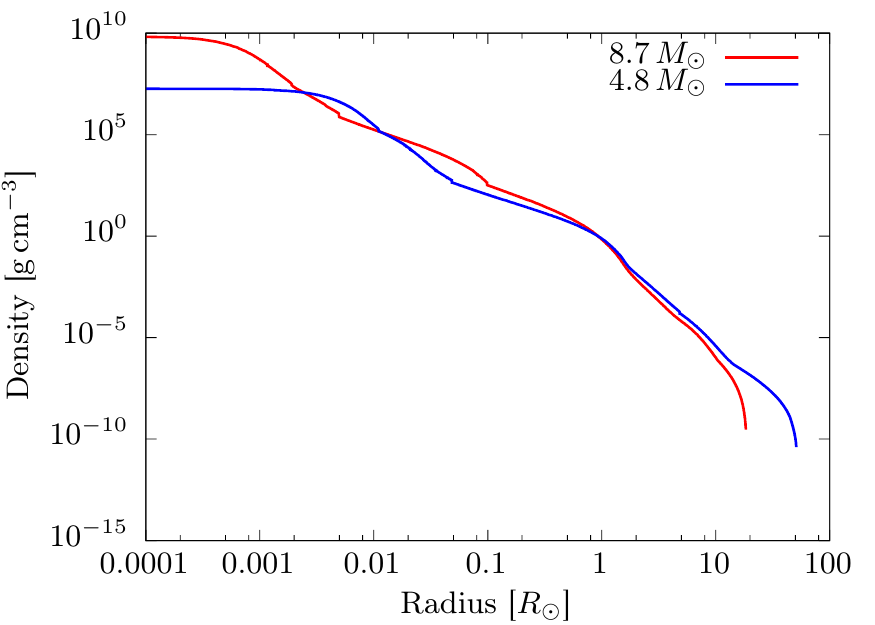}
 \caption{Density structures of the progenitors we used in our explosion simulations.\label{fig:progenitors}}
\end{figure}

The progenitor models are created using the stellar evolution code \texttt{MESA} \citep[v10108;][]{MESA1,MESA2,MESA3,MESA4}. We first create stars with masses of 30 and 16\,$\msun$ at the zero-age main sequence. The effect of ECI is expected to be largest in close binaries where the progenitors would inevitably have experienced mass transfer to their companion stars. To reproduce stellar models that will represent stars that have experienced mass transfer, we simply apply a certain mass loss rate of $10^{-3}\,\msun$ yr$^{-1}$ when the star enters the Hertzsprung gap phase. This should only roughly mimic the evolution of a star in a binary transferring its mass via Roche lobe overflow as we are not focusing on this phase. We switch off the mass loss when only a tiny fraction of hydrogen is left ($\lesssim0.05\,\msun$). The evolution is then continued up to the onset of CC for the larger progenitor. For the smaller progenitor we only follow up to Ne burning due to numerical diffulties in modelling the later burning stages. However, the structure of the envelope will almost be completely unaffected after this stage because the star will collapse in about a day. Therefore we use this stellar model as the pre-SN structure of the progenitor. By this time there is only a tiny amount of hydrogen left at the surface layers ($\lesssim0.03\,\msun$). Density structures of the progenitors we created are shown in Figure \ref{fig:progenitors}. Both have helium cores up to $\sim1\,\rsun$ and a dilute hydrogen envelope extending further out. Note that the nominal radii of these stars are larger than the binary separations we assume later (see section \ref{sec:step2}) in some cases. A real progenitor that has undergone binary evolution will have a much smaller radius depending on the orbital separation. Irrespective of the actual radii of the hydrogen envelopes in our models, however, we consider that the structures inside the helium cores and the subsequent explosions will be unaffected by the presence of these tenuous hydrogen envelopes with such small masses. We are therefore not concerned with the hydrogen envelope and the apparent contradiction between the stellar radii and the binary separations in our progenitor models. It can be considered that our progenitor models and the subsequent ejecta profiles are equivalent to that of helium stars with radii of $\sim1\rsun$ and will explode as type Ib or type IIb SNe depending on the amount of hydrogen left on the surface.

We then artificially explode these stars on the hydrodynamical code with the ``thermal bomb'' technique \citep[]{you07}. We first place the progenitor model with a mass $M_\mathrm{prog}$ on a one-dimensional spherical grid. Since mesh-based codes cannot treat vacuum, a dilute atmosphere is placed around the star which will not affect the dynamics. The star and atmosphere is covered with 1600 radial grid points where the innermost cell size is $2\times10^9$\,cm and increased in a geometrical progression outwards up to the outer boundary at $1.5\times10^{14}$\,cm. The star is covered in 365 and 638 cells for the 8.7 and 4.8\,$\msun$ progenitors respectively. For this step we do not use the hyperbolic self-gravity solver but simply calculate the gravitational force from the enclosed mass. We excise the central $M_\mathrm{PNS}=1.6\,\msun$\footnote{This is heavier than the typically observed masses of NSs, but it should be noted that this is the baryonic mass and the final gravitational mass will be smaller by $\sim10\%$, depending on the equation of state \citep[]{bom96}.} and set a reflective inner boundary condition to represent the surface of a proto-NS. We set an outgoing condition for the outer boundary. Then we inject an energy of $E_\mathrm{in}=E_\mathrm{exp}-E_\mathrm{env}$ to the inner few cells where $E_\mathrm{exp}$ is the explosion energy and $E_\mathrm{env}$ is the binding energy of the envelope. This will initiate an artificial explosion, where the final ejecta will have a kinetic energy of $E_\mathrm{exp}$. The ejecta mass will be $M_\mathrm{ej}=M_\mathrm{prog}-M_\mathrm{PNS}$. Although it is known that multidimensional hydrodynamical effects are important for the explosion mechanism of CCSNe, the energy gain region is small enough compared to the whole star and our 1D approximation will be sufficient to study the kinematics of the ejecta. We carry out four simulations with different explosion parameters, which we summarize in Table \ref{tab:explosion} along with part of the results. As well as the two different progenitor masses, we apply two different explosion energies, one with a canonical value ($E_\mathrm{exp}=10^{51}$\,erg) and one with a hypernova class energy ($E_\mathrm{exp}=10^{52}$\,erg). During the simulation, we record the time evolution of the physical variables at a fixed radius $r=R_f$ far out from the surface of the progenitor, which we use for the second step.

\begin{table}[t]
 \begin{center}
  \caption{Parameters and results for the explosion simulation.\label{tab:explosion}}
  \begin{tabular}{ccccc}
   \tableline\tableline
   Model & $M_\mathrm{prog}[\msun]$ & $E_\mathrm{exp}[10^{51}$\,erg]& $M_\mathrm{ej}[\msun]$ & $p_\mathrm{tot}$[cgs]\\
   \tableline
   7.1c & 8.7 &  1 & 7.1 & $8.0\times10^{42}$\\
   7.1h & 8.7 & 10 & 7.1 & $1.6\times10^{43}$\\
   3.2c & 4.8 &  1 & 3.2 & $3.3\times10^{42}$\\
   3.2h & 4.8 & 10 & 3.2 & $1.0\times10^{43}$\\
   \tableline
  \end{tabular}
 \end{center}
\end{table}

\subsection{Step 2: Collision of the SN Ejecta with the Companion}\label{sec:step2}
In the second step we simulate the phase where the SN ejecta of the primary collide with the companion star. Secondary stars at this point would usually still be on the main sequence (MS) unless they had very similar initial masses to the primary. In close binaries they will have experienced mass accretion from the primary star. It is known that post-mass accretion stars have slightly different structures from normal MS stars of the same mass \citep[e.g.][]{bra95}, but here we ignore this effect and use normal MS stars for the companion models. We create MS models using \texttt{MESA} with various masses, and various ages because stars grow in radii and thus change their structure even during the MS. Metallicity is fixed to the solar value (Z=0.02). The masses are taken to be $M_2=10$--20$\,\msun$ in order to ensure that the secondary star is heavy enough to lead to CC. MS stars in this mass range have radii in the range $R_2=5$--$9$\,$\rsun$ depending on the mass and age. For each companion star model, we simulate ECI at four different orbital separations $a=20,\,30,\,40,\,60\,\rsun$\footnote{In some models with the smallest separations the companion can be overfilling its Roche lobe. Such systems will not stably exist in nature, but we carry out the simulations nevertheless to understand the physics, in particular the systematics of ECI.}. The binary parameters used in our simulations are all listed in Table \ref{tab:results} along with the results.

In our simulations we ignore the effects of stellar rotation and orbital motions. The rotational deformation of stars will be small unless the spin is close to break-up, and stars will usually be tidally locked to the orbit for close binaries. The orbital motion is also negligible as it is much slower than the velocity of the ejecta. Hence we assume axisymmetry along the axis connecting the centers of the two stars and carry out 2D simulations.

The stellar models are placed at the origin of a 2D cylindrical grid. The computational domain extends from $r=0$ to $r=2.5\times10^{12}$\,cm in the radial direction and from $z=-1.25\times10^{12}$\,cm to $z=2.5\times10^{12}$\,cm in the longitudinal direction. One end of the longitudinal direction is taken shorter so that the exploding star does not enter the computational region. Other outer ends are placed at approximately 4--6 times the stellar radius, and we divide this domain into $(N_r\times N_z)=(600\times900)$ equally-spaced cells. Data from the first step is then mapped from the short end of the domain as an outer boundary condition, assuming that the ejecta follow homologous expansion. 

We use the hyperbolic self-gravity solver for self-gravity in this step. See \citet{RH16} for details of this solver. We set the gravitation propagation parameter to $k_g=5$, and apply the Robin boundary condition for the outer boundaries \citep[]{gus98}. The computational domain for the gravitational potential is taken approximately two times larger in every direction (the source term is set to 0 in the extended region) to reduce the magnitude of errors arising from the boundaries. We do not include the gravitational pull from the remaining NS of the primary in order to single out the pure hydrodynamical effects of ECI from the gravitational effect.

Simulations are followed up to the point where the bulk of the ejecta has finished interacting with the companion. We define the cells which have a negative total energy $\frac{1}{2}v^2+\epsilon+\phi<0$ as bound, where $v$ is the velocity, $\epsilon$ is the specific internal energy and $\phi$ is the gravitational potential energy. At each time step we record the total bound mass which is evaluated by integrating the mass contained in all the bound cells. We also record the total momentum contained in the bound cells to obtain the final impact velocity achieved by ECI.

\section{Results}\label{sec:results}
In the following subsections we review the results of the hydrodynamical simulations in each step.

\subsection{Explosion of the Primary}
\begin{figure}
 \centering
 \includegraphics[]{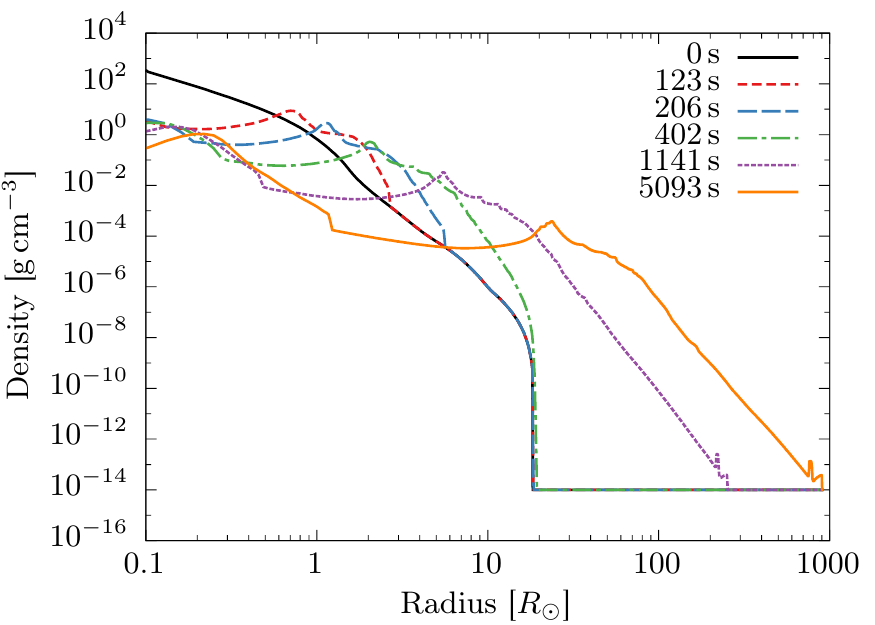}
 \caption{Radial density distributions at various times for the simulation 7.1c. Times are measured from the start of the simulation.\label{fig:explosion}}
\end{figure}

\begin{figure}
 \centering
 \includegraphics[]{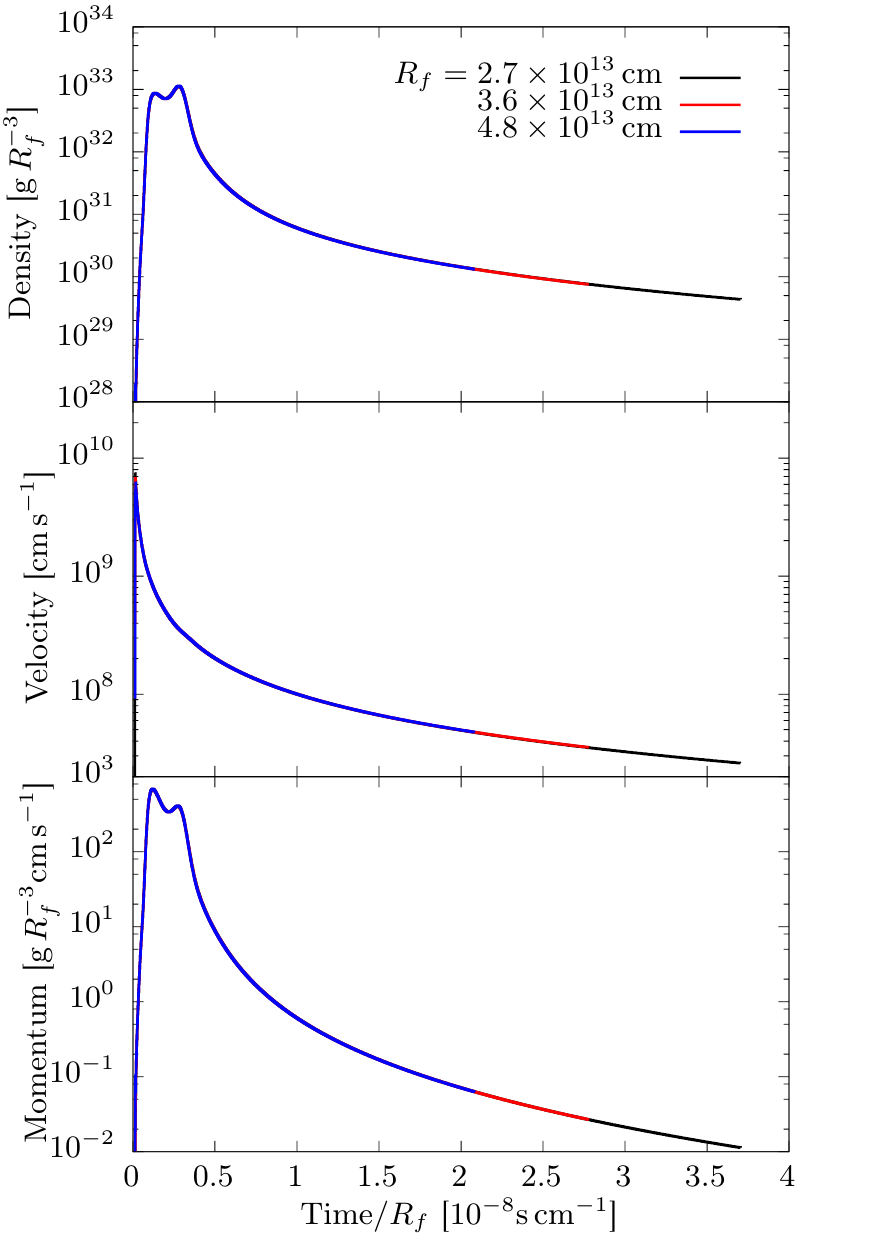}
 \caption{Evolution of various physical variables at different fixed radii outside the star.\label{fig:homologous}}
\end{figure}

As a representative model, we choose the 7.1c model to describe the dynamics of the explosion. Soon after we initiate the simulation, a shock wave is formed around the inner cells with the injected energy. Figure \ref{fig:explosion} shows the time evolution of the shock wave propagating through and penetrating the surface of the star. The shock accelerates as it runs down the density gradient and reaches the surface in $\sim400$\,s. After shock break out, the ejecta clearly follow a homologous expansion during the whole course of our simulation. This can be observed in Figure \ref{fig:homologous} where we plot the time evolution of various physical values observed at a fixed point $r=R_f$. Note that time has been rescaled by $R_f$ and density is also shown in units normalized by $R_f$. All three curves almost completely agree with each other after being rescaled, which means it is expanding self-similarly. This self-similarity is useful to be able to rescale ejecta profiles at arbitrary distances in the second step. We use the profile at $R_f=2.7\times10^{13}$\,cm for the second step since it traces deepest into the ejecta, but the choice of $R_f$ does not affect the results at all. It can also be seen that the ejecta expand as a dense shell, and the inner slower ejecta are small in mass. All other models that we simulated followed similar dynamics.

In the last column of Table \ref{tab:explosion} we list the total outgoing momentum $p_\mathrm{tot}$ integrated over the entire SN shell. It can be seen that the momentum is not related to the explosion properties in a simple manner.

\subsection{Collision of SN ejecta with the Companion}
\begin{figure*}
 \centering
 \includegraphics[]{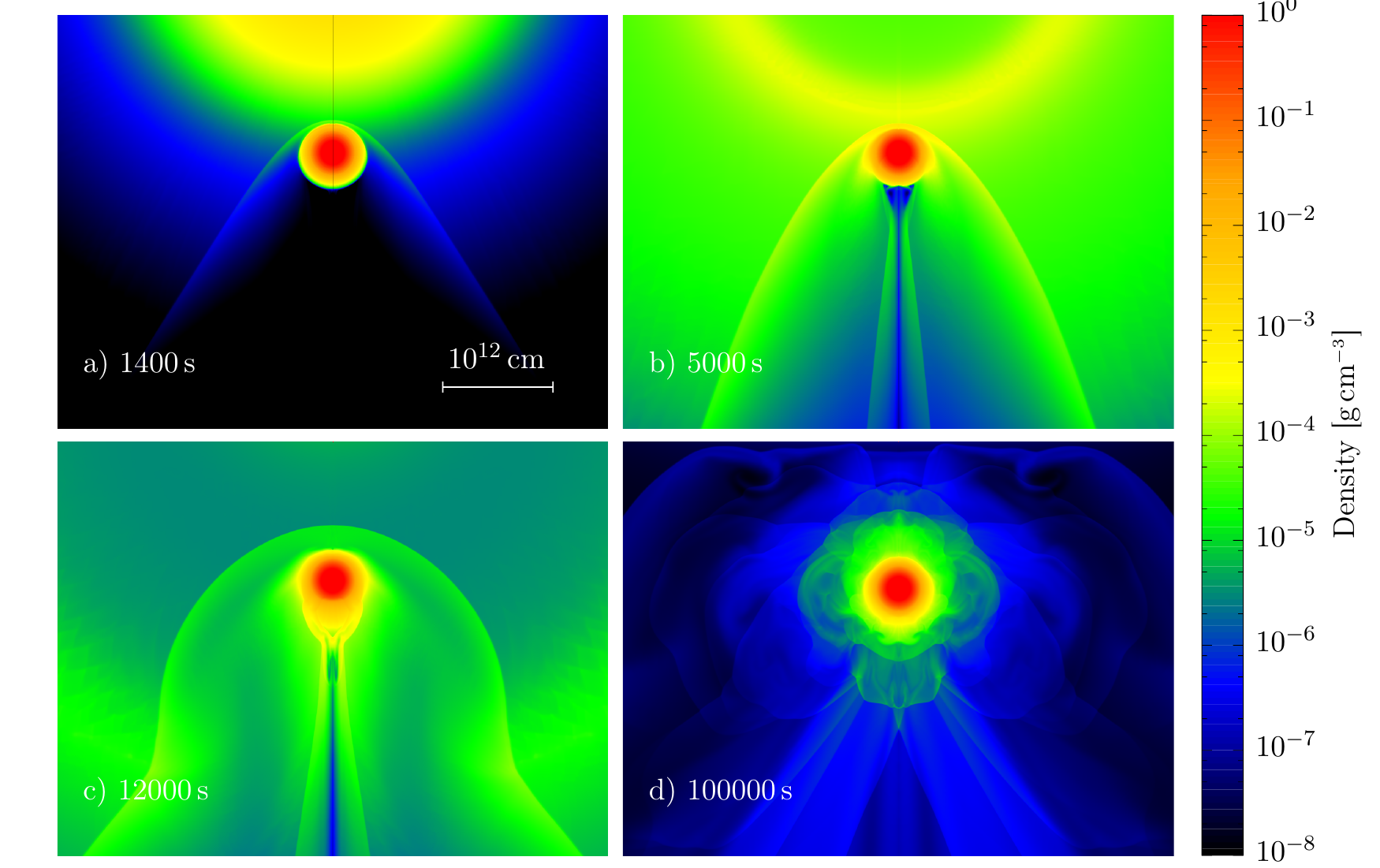}
 \caption{Snapshots of the density distribution in our ECI simulation with the M10R5MSa30-7c model. The time labelled on each panel show the time since the explosion.\label{fig:ECIdynamics}}
\end{figure*}

Here we first describe the dynamics of the whole ECI process. In Figure \ref{fig:ECIdynamics} we illustrate several snapshots of the density distribution in the simulation with the M10R5MSa30-7c model. As soon as the forefront of the ejecta reach the companion, it creates a bow shock in front of the star (panel a). The bow shock stays at the same position while the dense shell of the ejecta flow past (panel b and c). At the same time, a forward shock propagates through the star. The shock weakens as it ascends the density gradient on the front hemisphere but then regains its strength as it rushes down the density gradient on the other side of the star. Once it reaches the back end, it unbinds a small chunk of envelope matter. After the dense shell has finished flowing past, the heated up surface layers start to expand almost spherically (panel d). Because the bulk momentum of the ejecta is carried by the dense shell, no more momentum is added and the star carries on a uniform motion along the axis.

By this time the bulk momentum of the ejecta have also finished flowing past, so the star carries on a uniform motion along the axis.

\begin{figure}
 \centering
 \includegraphics[]{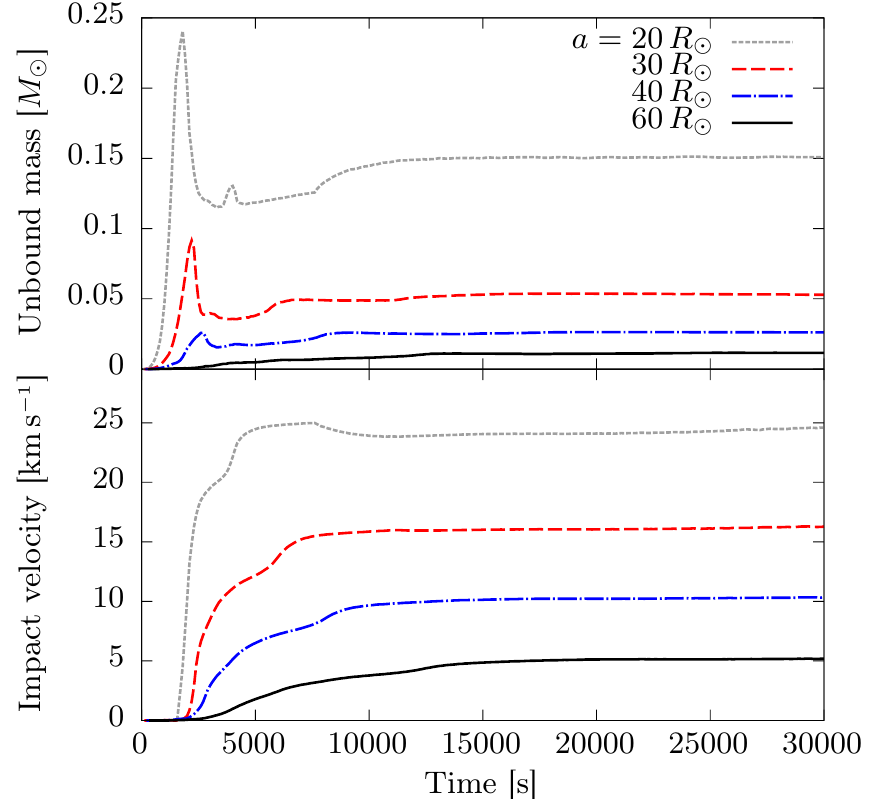}
 \caption{Time evolution of the unbound mass and impact velocity obtained from our simulations with ($M_2,R_2$)=($10\,\msun,8\,\rsun$) for the companion star and the 7.1c model for the explosion.\label{fig:timeevo}}
\end{figure}

Throughout our simulation we have recorded the mass of the total bound matter. Figure \ref{fig:timeevo} displays an example of the time evolution of the total bound mass along with the impact velocity. Impact velocities are calculated by dividing the total momentum contained in the bound cells by the total bound mass. In the simulations with the smaller orbital separations ($a=20,30,40\,\rsun$), there is a rapid rise in the unbound mass ($\sim2000$\,s) but then rapidly declines straight after. This is probably because the shock driven into the star is strong, so the internal energy behind the shock will be large enough to unbind matter. But the shock rapidly weakens as it travels deeper into the star, and the post-shock region can become bound again. The wider separation models do not show this initial spike because the shock is weaker. At a later time ($\sim7500$\,s), there is an extra amount of unbinding when the forward shock reaches the other side of the star and a chunk of surface matter is ablated off. All models have reached a steady value after $\sim15000$\,s. The impact velocity evolves in a more straightforward way. It rises proportionally to the incident momentum and reaches a steady value as soon as the ejecta shell has flowed past.

\begin{figure}
 \centering
 \includegraphics[width=0.48\textwidth]{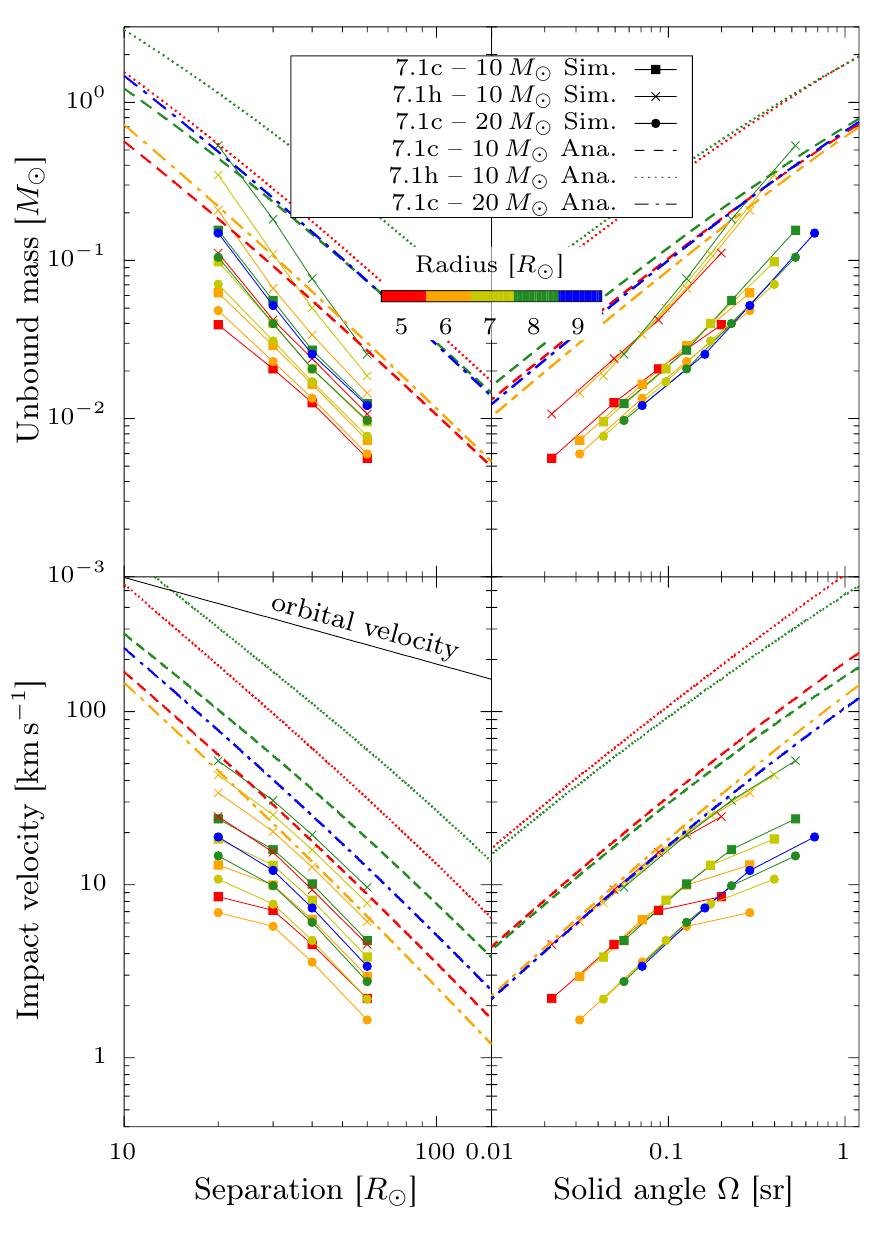}
 \caption{Final unbound mass (upper panels) and impact velocity (lower panels) obtained in our simulations with $M_\mathrm{ej}=7.1\,\msun$. Shapes of the plots discriminate the models with different explosion energy and companion mass, as listed in the legend. Colours of the plots indicate the different stellar radii of the MS star models used. Points with the same companion star model are connected with lines. Dashed, dotted and dot-dashed curves show the analytical estimates using the method described in \citet{whe75} for various parameters. The solid black curve in the lower left panel shows the orbital velocity of a $10\,\msun$ star in orbit with a $8.7\,\msun$ star.\label{fig:summary8.7}}
\end{figure}

\begin{figure}
 \centering
 \includegraphics[width=0.48\textwidth]{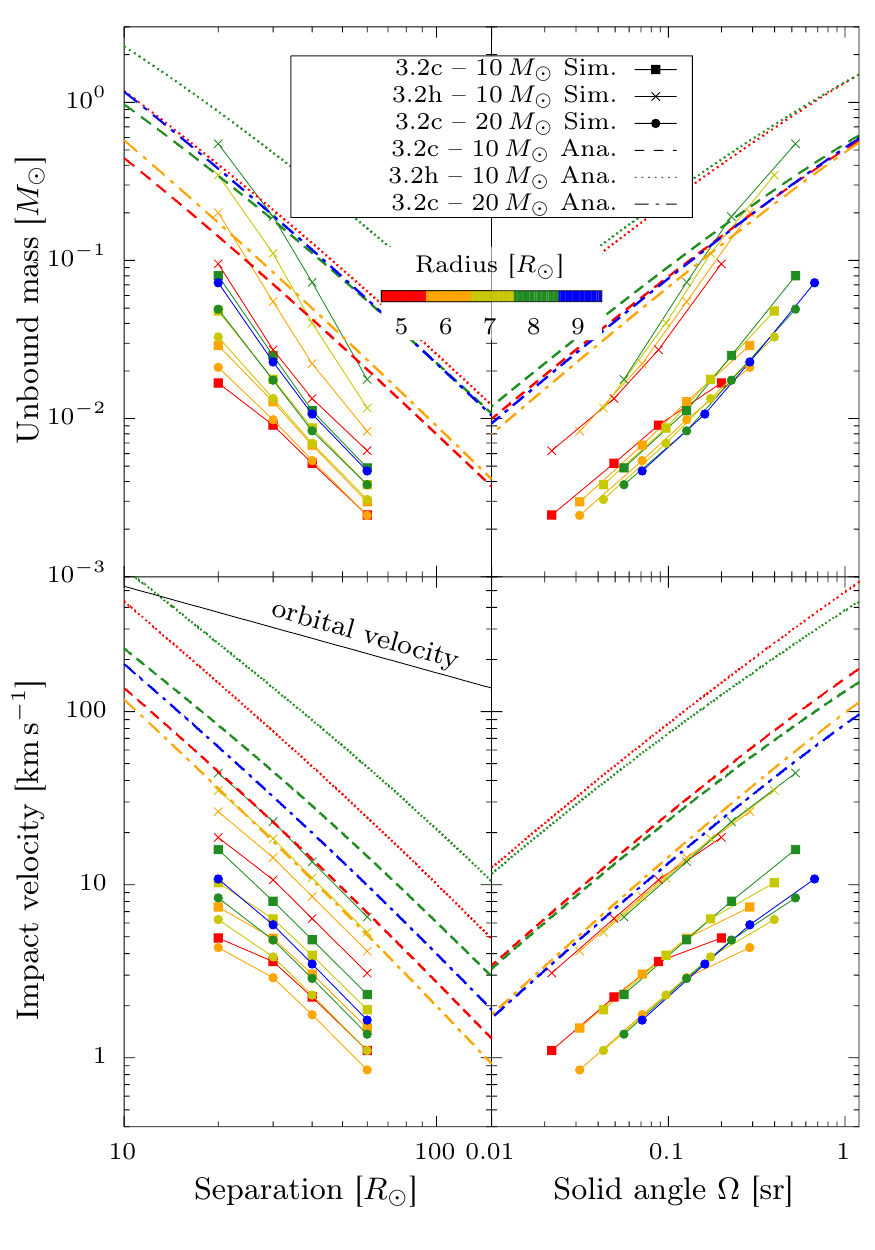}
 \caption{Same as Figure \ref{fig:summary8.7} but with $M_\mathrm{ej}=3.2\,\msun$.\label{fig:summary4.8}}
\end{figure}

In Figures \ref{fig:summary8.7} and \ref{fig:summary4.8} we display the final unbound mass and impact velocity obtained from some of our simulations. The full collection of results can be found in Table \ref{tab:results} along with estimates from the analytical model by \citet{whe75} and the momentum transfer efficiency $\eta$ that will be explained in section \ref{sec:momentum_transfer}. For each given companion star model, the unbound mass and impact velocity declines as separation increases. The dependence on separation roughly obeys a power law with slightly different powers for each stellar model, which is consistent with previous studies \citep[]{RH14,liu15b,rim16}. The dependence on stellar structure becomes clearer when plotted against the intersected solid angle $\Omega=2\pi\left(1-\sqrt{1-(R_2/a)^2}\right)$ as shown in the right panels. While the lines have a wide scatter on the left panels, all plots lie on a single line for a given explosion energy and companion mass in the right panels. This feature is more notable for the impact velocity, where the values lie on a straight linear trend. Only the $a=20\,\rsun$ models lie outside this relation, but some of these close systems already overfill their Roche lobes and are unlikely to stably exist in nature. It should be noted that this deviation from the power law occurs at a fixed separation and not at a fixed solid angle. This suggests that the impact velocity is not simply determined by the intercepted momentum but the absolute ram pressure of the ejecta plays a key role too.

In the figures we have also displayed the estimated values of unbound mass and impact velocity using the analytical models by \citet{whe75}. The analytical model overestimates the unbound mass by factors of $\lesssim12$, and the impact velocity by factors of $\sim4$--7 in comparison to our simulated results. The slopes also do not agree for the unbound mass, especially for the high energy explosion models where the simulations show a steeper dependence on separation. This may be due to the numerical resolution, which will be discussed in the next section. On the other hand, the slopes for the impact velocity agree quite well.

By comparing Figures \ref{fig:summary8.7} and \ref{fig:summary4.8}, it can be seen that the unbound masses and impact velocities are lower for explosion models with lower ejecta masses even with the same explosion energy. The differences in impact velocity is roughly directly proportional to the total outgoing momentum of the ejecta (see Table \ref{tab:explosion}). The differences in unbound mass also roughly proportional to the difference in ejecta masses, which was also seen in \citet{RH14}.

\subsection{Convergence and Consistency Tests}\label{sec:convergence}
To assess how firm our results are, we have performed a convergence test by varying the numerical resolution. We use the M10R8MSa30-7c model for the test. The simulation is repeated in four different resolutions ranging from $N_r\times N_z=400\times600$ to $1000\times1500$, where $N_r$ and $N_z$ are the number of gridpoints in the radial and longitudinal directions respectively. The ratio of $N_r$ and $N_z$ are fixed to ensure that the cells are equally spaced in both directions.

\begin{figure}
 \centering
 \includegraphics[]{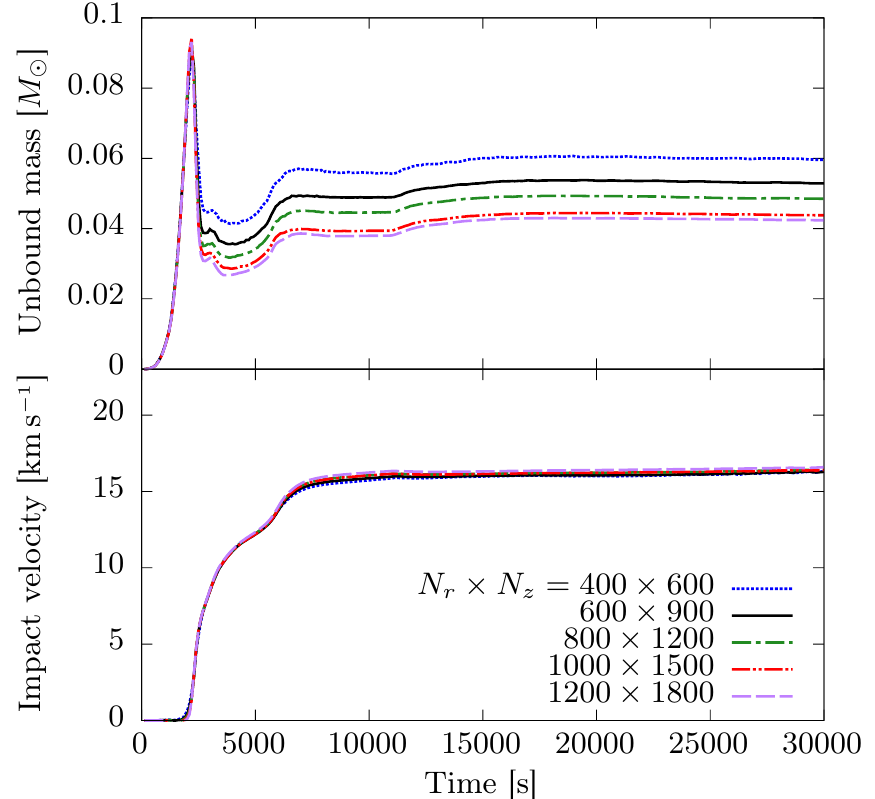}
 \caption{Time evolution of unbound mass and impact velocity for our numerical convergence test simulations.\label{fig:conv_test}}
\end{figure}

We display the time evolution of the unbound mass and impact velocity in Figure \ref{fig:conv_test}. From the upper panel it can be seen that the unbound mass decreases as the numerical resolution is increased. This was also observed in studies carried out with SPH codes \citep[]{liu15b,rim16}. Unfortunately, the results do not reach convergence even at the highest resolutions we have tested. This is probably because the amount of unbound mass is very small compared to the total stellar mass and is sensitive to how well the forward shock is resolved since most of the mass is unbound by ablation. With the standard resolution we used (black line), the removed mass is overestimated by a factor $\sim1.3$ judging from the converging trend. On the other hand, the impact velocity is less sensitive to the numerical resolution and has already reached convergence even with our lowest resolution. Thus, the results shown in Figures \ref{fig:summary8.7} and \ref{fig:summary4.8} for impact velocity are very firm, while we should regard the unbound masses as upper limits. 

To justify our usage of the hyperbolic self-gravity solver, we have carried out simulations with the normal Poisson solver for some of the models. When comparing the time evolution of the removed masses and impact velocities for the hyperbolic self-gravity and Poisson solver runs, the results agree within $\lesssim10^{-5}$ and $\lesssim10^{-3}$ respectively. 

\section{Discussion}\label{sec:discussion}

\subsection{Efficiency of Momentum Transfer}\label{sec:momentum_transfer}
It is evident from our results that the final impact velocity is smaller than the analytical estimates by \citet{whe75} by a factor of $\sim4-7$. This is partly because in their model the origin of the impact velocity is the reaction of the star by the ablation of surface matter, which is estimated by the injected momentum. This imparts roughly twice the incident momentum because all the ablated material is assumed to move opposite to the impact direction. However, the results disagree by more than a factor of two, implying that the incident momentum is not fully transferred to the companion.

\begin{figure}
 \centering
 \includegraphics[]{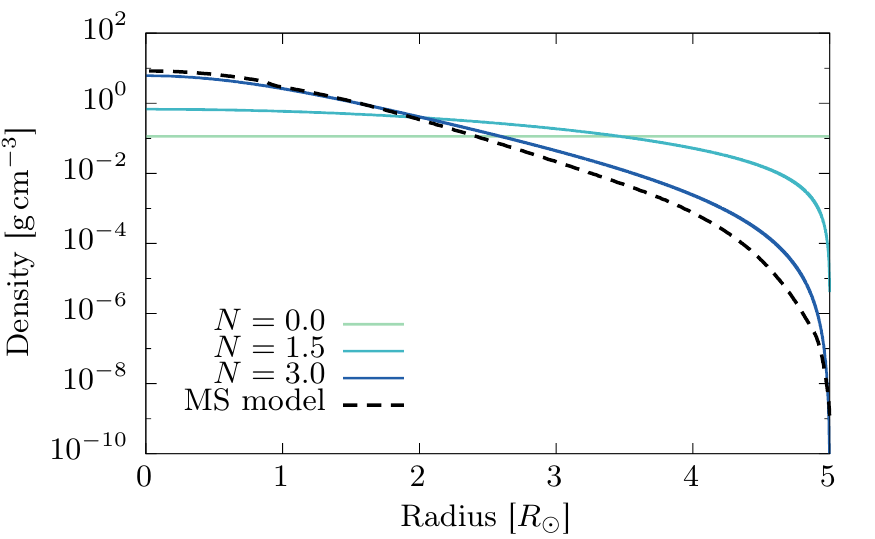}
 \caption{Comparison of density structures of the $(M_2,R_2)=(10\,\msun,5\,\rsun)$ stellar models used in our simulations.\label{fig:stellarstructure}}
\end{figure}

\begin{figure}
 \centering
 \includegraphics[]{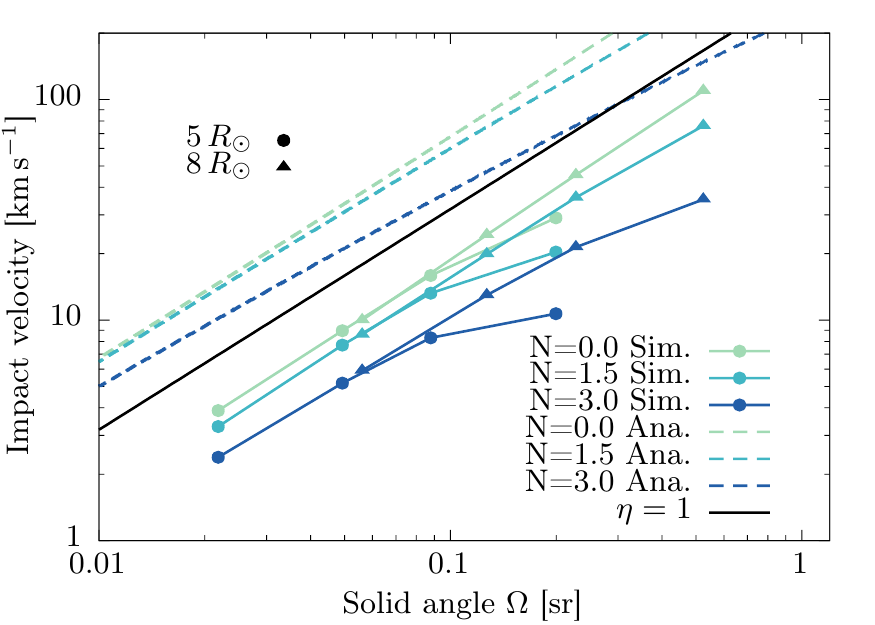}
 \caption{Impact velocities in our simulations with polytrope spheres with various polytropic indices and radii. The mass is fixed to $M_2=10\,\msun$ and the explosion model is 7.1c. Circles are results for $5\,\rsun$ stars and triangles for $8\,\rsun$ stars. Dashed lines show the analytical estimates using the model in \citet{whe75}. The solid black line shows a simple estimate assuming fully efficient momentum transfer from the incident ejecta ($\eta=1$).\label{fig:impact_pol}}
\end{figure}

\begin{figure}
 \centering
 \includegraphics[]{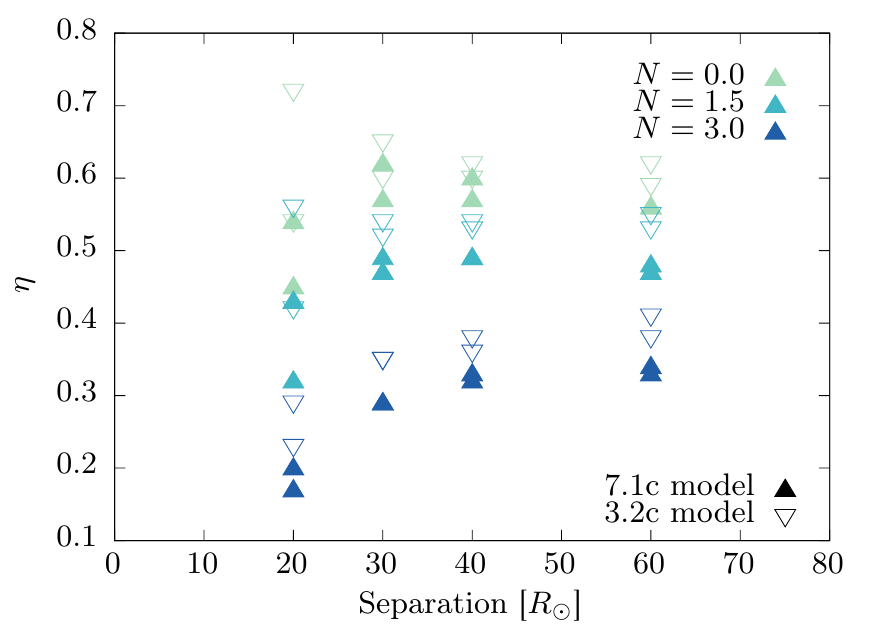}
 \caption{Values of $\eta$ obtained from the ECI simulations with polytrope companions. Colours of the plots denote the polytrope index of the companion and shapes of the plots denote the explosion model used in the simulation.\label{fig:eta_distribution}}
\end{figure}

In order to clarify how the impact velocity depends on structure, we carried out additional sets of simulations with various polytropes as companions. Figure \ref{fig:stellarstructure} displays a comparison of the density structures of the MS star model and polytrope spheres with various polytropic indices $N$. The MS star model has a very similar structure as the $N=3$ polytrope. In Figure \ref{fig:impact_pol} we show our results for some of the simulations with $N=0, 1.5, 3$ polytropes as companions. Although the companion mass and explosion properties are fixed, the impact velocity varies depending on the structure of the star, i.e. polytropic index. The solid black line shows a simple estimate for the impact velocity assuming no mass stripping and fully efficient momentum transfer. It can be seen that the simulated results are all below this line and the efficiency of the momentum transfer decreases as the polytropic index goes up. Here we will define the efficiency as
\begin{eqnarray}
 \eta\equiv \frac{M_{2,\mathrm{rem}}v_\mathrm{im}}{p_\mathrm{tot}\tilde{\Omega}},
\end{eqnarray}
where $M_{2,\mathrm{rem}}=M_2-M_\mathrm{ub}$ is the remnant mass of the companion, $p_\mathrm{tot}$ is the total outgoing momentum of the ejecta and $\tilde{\Omega}\equiv\Omega/4\pi$ is the fractional intersected solid angle of the companion\footnote{This is equivalent to the $\eta$ parameter introduced in \citet{tau98}.}. $M_\mathrm{ub}$ was significantly smaller than $M_2$ in all of our simulations, so we can safely assume that $M_\mathrm{2,rem}\sim M_2$. Because all slopes were roughly parallel with the black line, it seems that $\eta$ depends strongly on the polytropic index. We have listed values of $\eta$ of all our polytrope runs in Table \ref{tab:polytroperesults}. We plotted all the $\eta$ values listed on Table \ref{tab:polytroperesults} in Figure \ref{fig:eta_distribution}. It is clear that the polytropic index has the largest influence on the value of $\eta$. For a fixed companion model, the simulations with smaller ejecta mass show larger values of $\eta$. There is also an increasing trend for the $N=3$ models whereas the $N=0.0, 1.5$ models show a weaker dependence on the separation. All these dependences indicate that the value of $\eta$ depends on both the incident ejecta and the companion star structure.

\begin{figure*}
 \centering
 \includegraphics[]{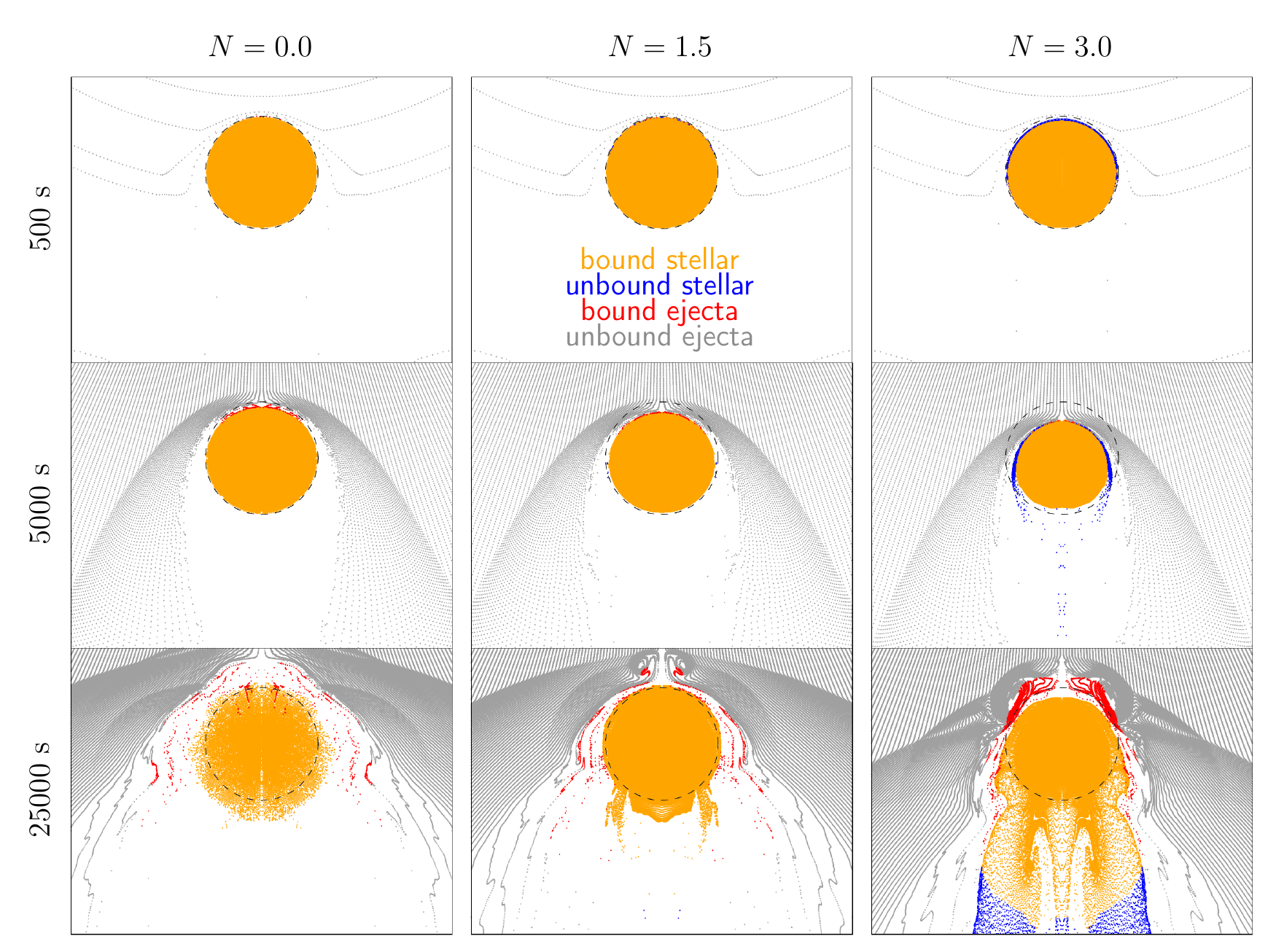}
 \caption{Snapshots of the positions of tracer particles in the simulations with polytrope models. The colours of the particles indicate whether they are bound stellar material (orange), unbound stellar material (blue), bound ejecta (red) or unbound ejecta (grey). The black dashed circle marks the initial position of the star with a radius of $5\,\rsun$.\label{fig:tracer_snapshots}}
\end{figure*}

To physically understand how the $\eta$ parameter is determined, we have carried out simulations with tracer particles that simply follow the motion of the fluid. Each particle carries the information of its origin (stellar or ejecta matter) and has a different mass. This allows us to better understand how the ejecta material interact with the companion star and whether it mixes with the stellar material. Figure \ref{fig:tracer_snapshots} displays various snapshots of the positions of the tracer particles in our simulations with the secondary star radius $R_2=5\,\rsun$ and the orbital separation $a=40\,\rsun$. The top panels show distributions at the onset of ECI, middle panels represent the main interaction phase and bottom panels a late time where most of the interaction has finished and the star is expanding. Each particle is coloured based on whether it originates from the star or ejecta, and whether it is bound to the star or not. In the left and middle panels ($N=0, 1.5$), there are almost no unbound stellar particles, indicating that the amount of mass removed due to ECI is small. For the $N=3$ model, the number of particles that have become unbound is larger, but the total unbound mass is comparable or even smaller than the other models. At the later stages there is a small amount of accretion of the slow ejecta matter onto the stellar surface (red particles). Some of the accreted matter is mixed deep into the interior of the envelope due to convective motions in the $N=0$ model. This contamination will be discussed later on. Most of the ejecta particles do not mix with the star but just move aside and flow around. The ejecta particles are imparting part of their momentum at this point, when they change direction to avoid mixing with the star.

\begin{figure}
 \centering
 \includegraphics[]{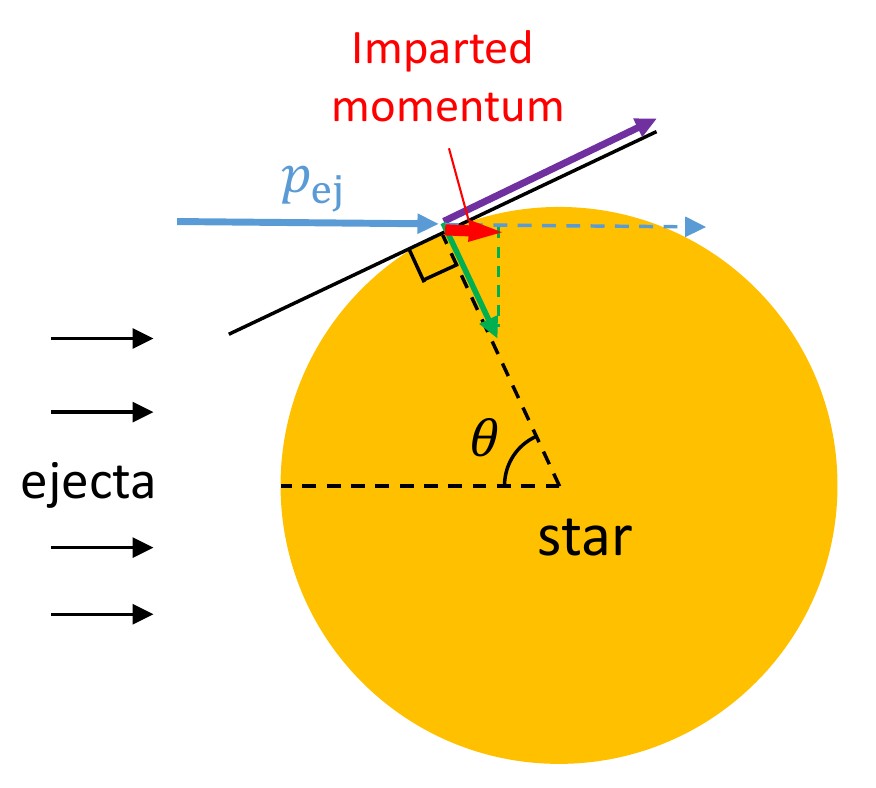}
 \caption{Schematic picture of our simple model of momentum transfer from the SN ejecta to the companion star.\label{fig:momentumtransfer}}
\end{figure}

We model this in a simple way that we schematically express in Figure \ref{fig:momentumtransfer}. The ejecta are assumed to hit a hard stellar surface in parallel rays from the left. We set a spherical coordinate system taking the origin at the centre of the companion and the axis pointing towards the exploding star. We shall first consider the momentum imparted by an ejecta mass element impacting the surface of the companion at an inclination angle $\theta$. The incoming momentum of this mass element (blue arrow) can be decomposed into components that are parallel (purple) and perpendicular (green) to the stellar surface. The ejecta need to impart this perpendicular component (or more) to the star in order to move away from the star and not merge with it. The green component can be further decomposed into components parallel and perpendicular to the symmetry axis, where the perpendicular components will cancel out due to axial symmetry. This will leave us with a net momentum pointing away from the exploding star with an amplitude $\cos^2\theta$ times the original incoming momentum (red arrow). By integrating the imparted momentum over the entire hemisphere facing the ejecta, we get an ideal efficiency
\begin{eqnarray}
 \eta_\mathrm{ideal}=\frac{\int^{2\pi}_0d\phi\int^\frac{\pi}{2}_0\sin\theta\cos^3\theta d\theta}{\int^{2\pi}_0d\phi\int^\frac{\pi}{2}_0\sin\theta\cos\theta d\theta}=\frac{1}{2},\label{eq:eta_ideal}
\end{eqnarray}
which roughly agrees with our simulated results for the $N=0$ polytrope ($\eta=0.6$). It should be noted that the efficiency can increase if the ejecta bounce off the star more violently, rather than sliding away parallel to the surface, which may explain the slight deviation between the model and the simulated results.

\begin{figure}
 \centering
 \includegraphics[]{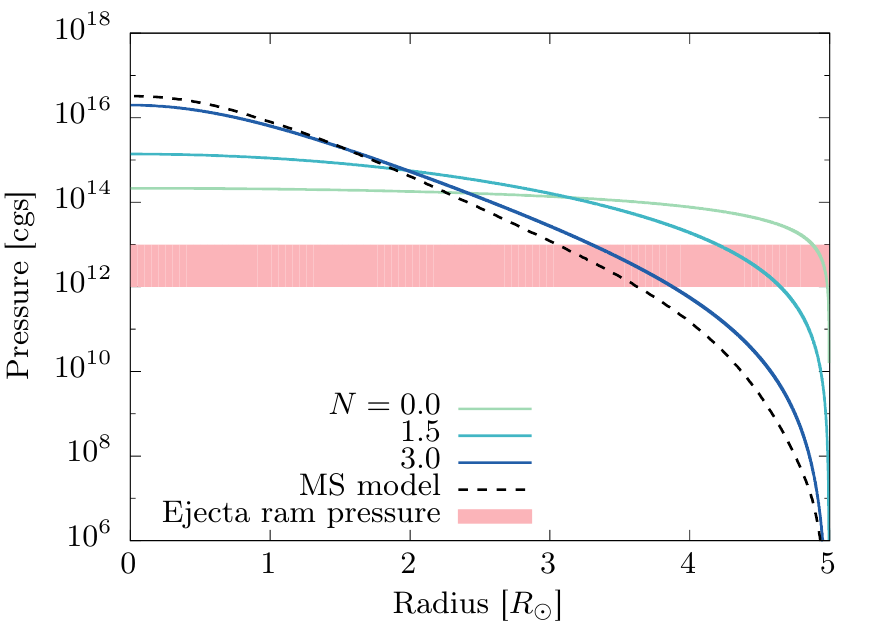}
 \caption{Pressure distributions inside polytrope spheres and the MS model with $(M_2,R_2)=(10\,\msun,5\,\rsun)$. The shaded region indicates the average ram pressure of the ejecta when the bulk of the matter flows past the star.\label{fig:pressurebalance}}
\end{figure}

All other stellar models with higher polytrope indices had lower values for $\eta$. Using tracer particles again helps us to understand this lower efficiency. As soon as the ejecta touch the surface of the star, they push away some of the surface matter as was assumed in the models of \citet{whe75}, but mainly compress the star into a smaller volume. This can be seen in the middle panels in Figure \ref{fig:tracer_snapshots} where the orange region is smaller than the dashed circle which marks the initial radius. The degree of the compression is roughly determined by the radius in which the pressure inside the star balances the ram pressure of the ejecta, which we show in Figure \ref{fig:pressurebalance}. This compression reduces the cross section of ECI, making way for more of the momentum to flow past the star without interacting. Thus the total momentum intersected by the star will also be reduced accordingly, leading to lower impact velocities. The differences in $\eta$ among the different models are proportional to the reduction in cross sectional solid angle. This argument can be extended to our MS models, where the radii at which the pressure balances the ram pressure are slightly smaller for the MS models than the $N=3$ polytropes. Values of $\eta$ listed in Table \ref{tab:results} are also systematically smaller than that of the $N=3$ polytrope runs. It also explains the shallower slopes for the more centrally concentrated stars.

Our model is also consistent with the dependences on ejecta models and companion mass. For example, the impact velocities are directly proportional to the total momentum in the ejecta for a fixed companion mass. This applies to all four ejecta models we have simulated, owing to the fact that the peak ram pressure in the ejecta models were similar within a factor $\lesssim4$. The similar ram pressures lead to similar compression factors, which implies that $\eta$ will be similar too. When varying the companion mass, a simple estimate will give an impact velocity inversely proportional to the mass. However, our simulations with $20\,\msun$ companions showed impact velocities slightly higher than halving the results for the $10\,\msun$ companions. This is due to the higher surface pressure for the higher mass models, leading to less compression, and increases the efficiency $\eta$.

Gathering up what we have understood from our simulations, the momentum transfer efficiency $\eta$ is not an exactly fixed quantity for each stellar model. In most models listed in Table \ref{tab:polytroperesults}, $\eta$ tends to increase as the orbital separation increases. We carried out one extreme case with $a=400\,\rsun$ (M10R5P3a400-3c model) where we find that the value of $\eta$ was significantly larger than the closer separation models. This verifies our model that $\eta$ depends on the balance between the pressure distribution inside the star and the ejecta ram pressure. The general form of $\eta$ would be
\begin{eqnarray}
 \eta_\mathrm{ana}=\frac{1}{2}\left(\frac{r_{p=p_\mathrm{ej}(a)}}{R_2}\right)^2, \label{eq:eta_anal}
\end{eqnarray}
where $r_{p=p_\mathrm{ej}(a)}$ is the radius in the star where the pressure equals the ram pressure of the incident ejecta. Here we have ignored the effect of mass stripping since the unbound masses were significantly smaller than the total companion masses.
However, it appears safe to assume that for MS stars, as listed in Table \ref{tab:results}, the efficiency is roughly $\eta\sim1/3$ for most ranges of separation due to the steep stellar pressure gradient compared to the variation in ejecta ram pressure, and for wider binaries the effect of ECI is negligible anyway. The value of $\eta\sim1/3$ is roughly consistent with the studies for lower mass companions \citep[]{rim16}.

Another interesting fact that can be derived from Table \ref{tab:polytroperesults} is that the unbound mass does not depend on the polytropic index monotonically. When compared among models with the same radii and explosion parameters, the $N=1.5$ models have larger unbound masses when $\Omega$ is small but the $N=0$ models rapidly increase as $\Omega$ increases. This should also be related to the steepness of the pressure or density gradient near the surface of the star, but we will leave to future work to understand what determines the unbound mass.

\subsection{Influence of the Impact Velocity on the Resulting Orbit}\label{sec:orbit}
All impact velocities obtained in our simulations were much smaller than the orbital velocity and average observed NS kick velocities. Thus it is unlikely that the impact velocity will have a large influence on the resulting orbit. Within the parameter range we have explored, ECI only reduced the survivability of the binary by $\lesssim0.5\,\%$.

\begin{figure}
 \centering
 \includegraphics[]{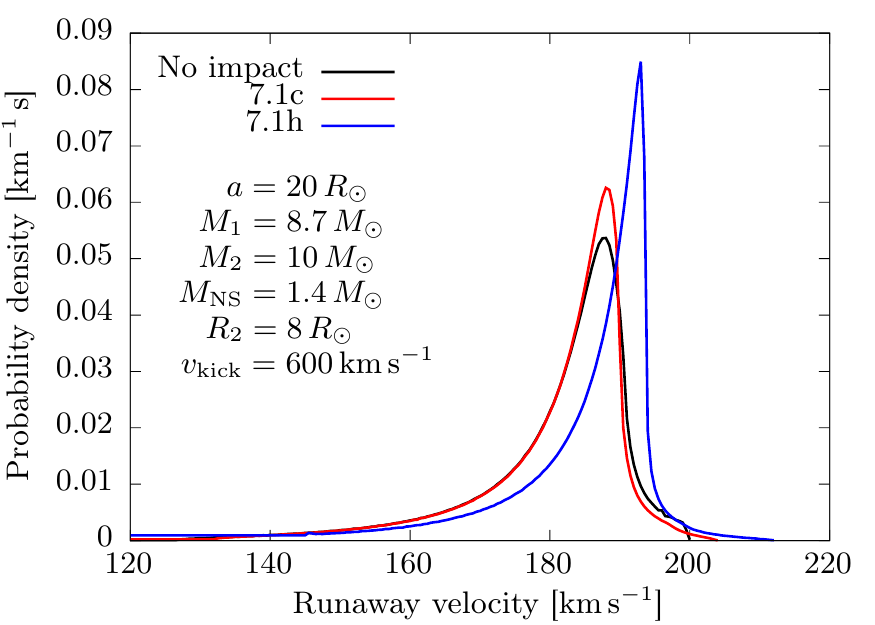}
 \caption{Probability distribution of the runaway velocity of the companion star after a SN explosion. Parameters used in the calculation are listed below the legend. Values for the impact velocity and removed mass were taken from our simulations with different explosion energies (red and blue) and are compared to the case assuming no impact (black).\label{fig:runawayvel}}
\end{figure}

 For the disrupted binaries, we calculated the runaway velocity of the companion star using the equations derived in \citet{tau98}. In Figure \ref{fig:runawayvel} we show the probability distribution of the runaway velocity of the companion in one particular case. In the calculation we have assumed the NS kick velocity to be $v_\mathrm{kick}=600$\,km\,s$^{-1}$ and randomly oriented from an isotropic distribution over the sphere. When assuming that the SN shell has no impact on the companion, the runaway velocity peaks at $\sim188$\,km\,s$^{-1}$ (black line). With an impact velocity $v_\mathrm{im}=25$\,km\,s$^{-1}$ taken from our 7.1c simulation, the distribution is just slightly distorted without changing the peak velocity (red line). With a higher impact velocity of $v_\mathrm{im}=55$\,km\,s$^{-1}$ taken from our 7.1h simulation, the whole distribution is shifted to the right, peaking at $\sim193$\,km\,s$^{-1}$ (blue line).

\begin{figure}
 \centering
 \includegraphics[]{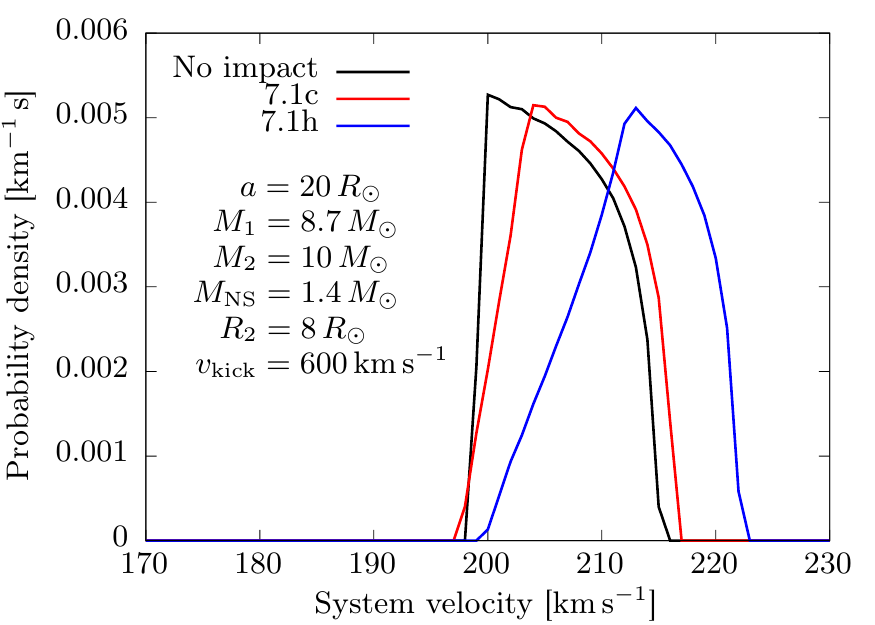}
 \caption{Probability distribution of the system velocity of surviving binaries after SN explosion. The same parameters have been used as in Figure \ref{fig:runawayvel}\label{fig:systemvel}}
\end{figure}

Surviving binaries can also acquire a velocity due to the sudden mass loss and the NS kick. Such velocities can drive the surviving binary out of the galaxy, and eventually produce binary NS systems at locations far from the galactic disk. This has been proposed as a possible explanation of the offsets of short gamma-ray bursts from their host galaxies \citep[]{bran95,fon13}. We have also calculated the influence of ECI impact velocities on the resulting system velocity, which we show in Figure \ref{fig:systemvel}. In the calculation we define the binary to have ``survived'' when the post-SN eccentricity of the orbit is $e<1$ and the post-SN periastron passage is $a_f(1-e)>R_2$. The second condition is required to exclude systems where the NS plunges into the companion star's envelope, which may lead to the formation of Thorne-\.Zytkow objects. This condition will be modified if we take into account the post-ECI inflation of the star and will result in fewer survivals. When we assume that there is no ECI, the system velocity peaks at the lower end and follows a bottom-heavy distribution. The impact velocity pushes the maximum velocity up and also alters the distribution to more top-heavy shapes. This may slightly extend the offsets of binary NS mergers from their host galaxies, which can be compared with future observations of short gamma ray bursts and kilonovae.

\subsection{Observational Signatures of ECI}\label{sec:observational}
There have been many attempts made to search for binary companions to CCSN progenitors \citep[e.g.][]{mau04,mau15,fol14,fol16,dyk16,ryd18}. In the successful cases where a companion was actually detected, it provides us with valuable information on the possible evolutionary paths of the progenitor towards the SN. To provide better constraints on the system it is important to understand how post-ECI companions look like and how various features relate to the pre-SN binary parameters. In this section we will discuss several possible observational features of post-ECI stars.

One of the consequences of ECI is some possible reddening of the companion as was predicted in many previous studies \citep[]{pod03,sha13,pan13,RH15}. As can be seen in Figure \ref{fig:tracer_snapshots}, the star can inflate due to the heat excess injected into the envelope. The expansion is too large and the time scale is too long to be fully followed in the hydrodynamic simulations. Here we will predict the long term evolution of the companion after ECI using \texttt{MESA} again. In this section we will focus on the binary models with the 7.1c model for the explosion and ($M_2,R_2$)=($10\,\msun,5\,\rsun$) model for the companion.

\begin{figure}
 \centering
 \includegraphics[]{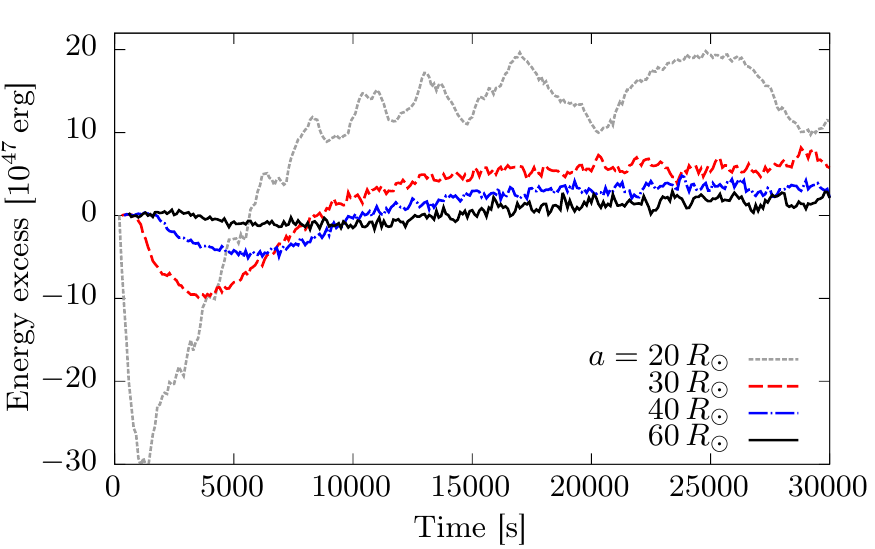}
 \caption{Time evolution of the energy excess of the companion star. Same binary models are used as in Figure \ref{fig:timeevo}.\label{fig:energytimeevo}}
\end{figure}

From the hydrodynamic simulations we can obtain the total amount of energy injected into the star. In Figure \ref{fig:energytimeevo} we show the time evolution of the energy excess obtained in our hydrodynamic simulations. The energy excess is calculated by integrating the total energy over all the bound cells and subtracting the initial binding energy. While the ejecta shell is passing through the star, the total energy decreases because of the compression of the envelope. The surface material is temporarily pushed deeper into the gravitational potential well, showing stronger binding energies. After the bulk of the ejecta have flown past, the envelope expands back outwards trying to retain hydrostatic equilibrium, and the binding energy decreases. It can be seen that most models reach a steady value after $\sim15000$\,s. The model with $a=20\,\rsun$ does not become steady, but rather fluctuates around a certain value. All models have higher energies at the end than before the explosion, the excess ranging from $1.0\times10^{47}$\,erg to $1.6\times10^{48}$\,erg. For example, the energy excess for the $a=30\,\rsun$ model is $\sim5.5\times10^{47}$\,erg whereas the intersected energy is $\sim7\times10^{48}$\,erg, implying an energy injection efficiency of $\sim8\,\%$. The rest of the energy is simply not transferred to the star, and partly taken away as the kinetic energy of the unbound matter. All other models also show similar efficiencies of $\sim8$--$10\,\%$. This can be explained in a similar picture as for the momentum transfer efficiency. In Figure \ref{fig:momentumtransfer}, the incoming ejecta fluid element is assumed to bounce away tangentially to the surface of the star and impart part of its momentum in doing so. At the same time it is imparting part of its kinetic energy to the stellar material which should quickly thermalize into heat. The imparted kinetic energy is determined by the component perpendicular to the stellar surface of the incoming ejecta velocity (green arrow in Figure \ref{fig:momentumtransfer}). In this model, if the incoming ejecta fluid element has a velocity $v_\mathrm{ej}$, the velocity is reduced to $v_\mathrm{ej}\sin\theta$ (purple arrow in Figure \ref{fig:momentumtransfer}) after being bounced off. This will reduce the kinetic energy of the ejecta fluid element by a factor of $\sin^2\theta$, meaning that the remaining factor of $\cos^2\theta$ of the energy should have been transferred to the star. By integrating this over the entire hemisphere facing the primary, we get an energy transfer efficiency $1/2$ from the same calculation as in Equation \ref{eq:eta_ideal}. This is the fraction of incident kinetic energy that is imparted to the star. Although the energy of the ejecta is initially dominated by kinetic energy, the bow shock in front of the companion star will compress the ejecta fluid element and convert part of the energy into thermal energy. By simply applying the Rankine-Hugoniot condition in the strong shock limit, the post-shock ejecta will have converted $2/(\gamma+1)$ of its kinetic energy into thermal energy where $\gamma$ is the adiabatic index. The flow around the star is too fast to transfer any heat, so the thermal energy component will not contribute to the energy injection. Thus the actual energy injection efficiency can be written as
\begin{eqnarray}
 \vartheta=\frac{1}{2}\frac{\gamma-1}{\gamma+1}\left(\frac{r_{p=p_\mathrm{ej}(a)}}{R_2}\right)^2=\frac{\gamma-1}{\gamma+1}\eta_\mathrm{ana}.\label{eq:energyinjection}
\end{eqnarray}
For a $\gamma=5/3$ gas this becomes $\vartheta=1/4\eta_\mathrm{ana}\sim1/12$, which is in good agreement with the simulated results ($\sim8$--$10\%$).

\begin{figure}
 \centering
 \includegraphics[]{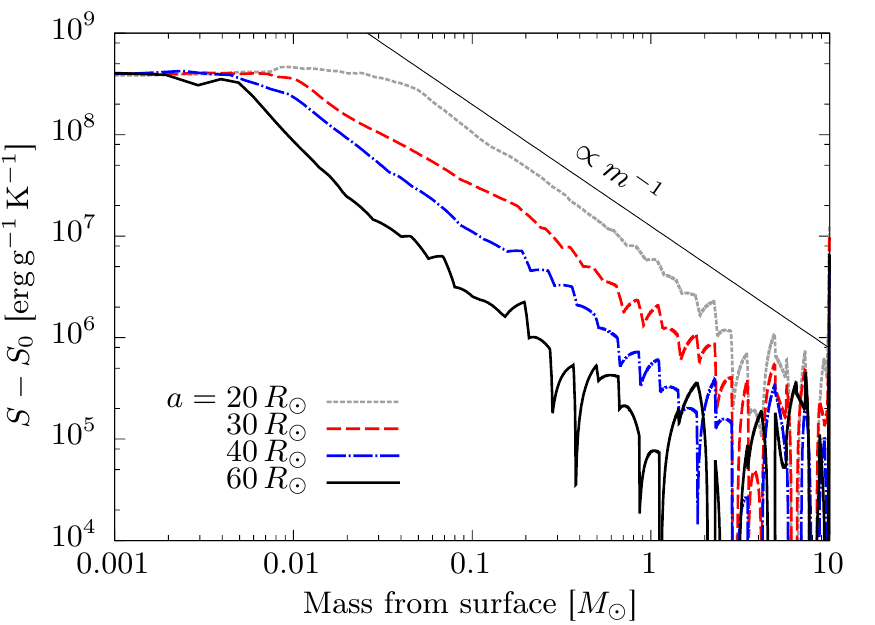}
 \caption{Distribution of entropy excess in the 1D reconstructed stellar models. Horizontal axis is taken as the mass coordinate from the surface. The black solid line shows a slope inverse proportional to the mass. Definitions of $S$ and $S_0$ are given in the text.\label{fig:reconstruction}}
\end{figure}

To identify where in the star the injected energy is deposited, we reconstruct our multidimensional results into a 1D distribution angle-averaged around the centre of mass. We then compare the final distribution of energy in the star with that of a star which has not experienced ECI. Figure \ref{fig:reconstruction} shows the distribution of the excess of entropy in the reconstructed stars, where $S$ and $S_0$ are the entropy from the reconstructed post-ECI and non-ECI models respectively. The reconstruction becomes less reliable towards the centre due to the small number of cells to average over. It is evident that the star is most heated near the surface, having a rather constant heating rate that starts to decrease inverse proportionally to the mass as it goes in. This is because the shock weakens as it climbs the density gradient in the star and deposits less energy in the high density regions. The radius at which the decline starts depends on the intersected mass, and the mass coordinate of this transition is roughly equivalent to half the intersected mass.

With the above information, we mimic the envelope heating by ECI on \texttt{MESA}. We take similar procedures to those of previous studies \citep[]{pod03,sha13,RH15}. First, we take the stellar model that was used for the initial condition in the hydrodynamical simulation and remove some surface material by applying a mass loss rate of $10^{-1}\,\msun$\,yr$^{-1}$. We stop the mass loss when the mass of the star reaches the post-ECI mass obtained from the simulation. Then we apply a heating rate
\begin{eqnarray}
 \dot{\epsilon}(m)=\frac{E_h}{\tau_hm_h}\cdot \frac{\min\left(1,m_h/m\right)}{1+\ln\left(M_{2,\mathrm{rem}}/m_h\right)},
\end{eqnarray}
where $m$ is the mass coordinate from the surface. This equation mimics the heat excess distribution which we obtained in Figure \ref{fig:reconstruction}. $E_h$ and $M_\mathrm{2,rem}$ are the total injected energy and post-ECI mass respectively, which we take from our simulation results. $\tau_h$ is the timescale of the heating, which we simply set as 1\,year. The results do not change so much with different choices of $\tau_h$ as long as it is shorter than the surface thermal timescale. $m_h$ represents the mass coordinate where the entropy excess starts to decrease, and we define it as $m_h=M_\mathrm{ej}(1-\sqrt{1-(R_2/a)^2})/4$ to match the heat distribution in Figure \ref{fig:reconstruction}. We switch off the heating after a time $\tau_h$, when the star has gained an energy excess of $E_h$. Then the star is left to evolve for $\sim10^4$\,years until it completely retains its original state.

\begin{figure}
 \centering
 \includegraphics[]{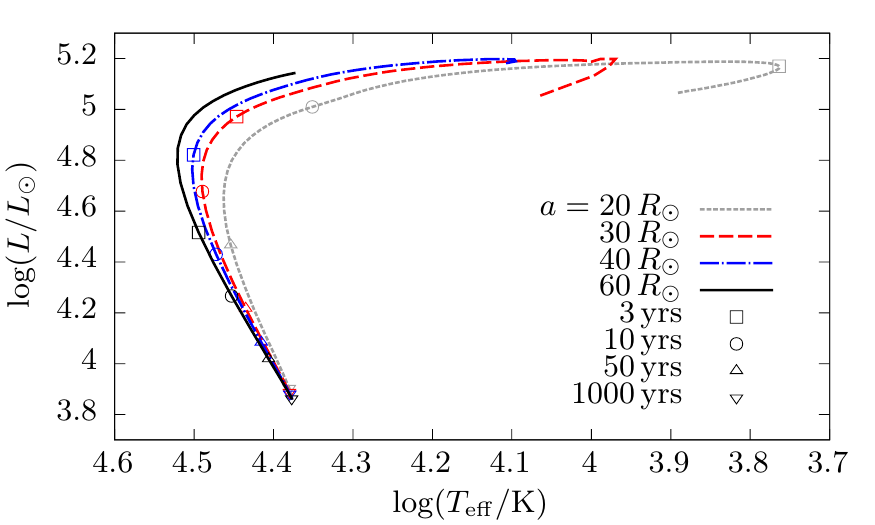}
 \caption{Evolution of the post-ECI stars in the HR diagram. Symbols mark the times elapsed since the heat was injected.\label{fig:HRdiagram}}
\end{figure}

\begin{figure}
 \centering
 \includegraphics[]{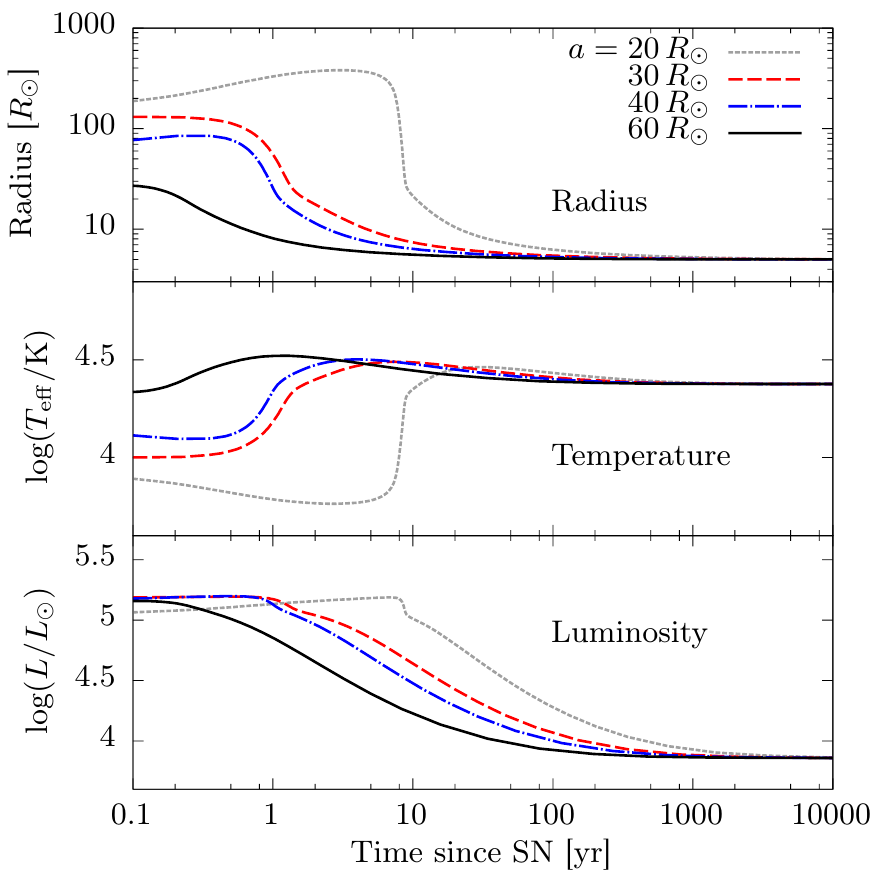}
 \caption{Time evolution of various stellar properties after ECI.\label{fig:heatevo}}
\end{figure}

Figure \ref{fig:HRdiagram} shows the evolution of the stars in the Hertzsprung-Russell (HR) diagram. In all the cases shown here, the luminosity increases by more than an order of magnitude straight after the heating. The models with closer orbits have energies injected into deeper layers, leading to a large expansion of the star and thus lower surface temperatures. However, this high luminosity only lasts for about a thermal timescale of the surface layer where most of the energy is deposited ($\sim m_h$). The star remains slightly overluminous while it radiates away the energy deposited deeper in the envelope, and after about $\sim1000$\,years it completely recovers to its original state. This can be better observed in Figure \ref{fig:heatevo} where we plot the time evolution of radius, temperature and luminosity of the companion after ECI. One possibly interesting feature is that the radius can expand dramatically and sometimes exceed the original orbital separation. The orbital separation would have changed by this time due to SN mass loss and the NS kick, but this expansion can still increase the possibility of the star engulfing the primary NS. It is not clear what would happen when a NS is captured in this thin inflated envelope, but there are possibilities of additional mass loss, orbital circularization, or formation of  Thorne-\.Zytkow objects \citep[]{tho77}. Even if the expansion is not large enough to engulf the primary NS, it can exceed its Roche lobe and initiate an extremely short term mass transfer phase. Although the total transferred mass may be small, it may be sufficient to create some planets around the primary NS \citep[]{nak91}.

\begin{figure}
 \centering
 \includegraphics[]{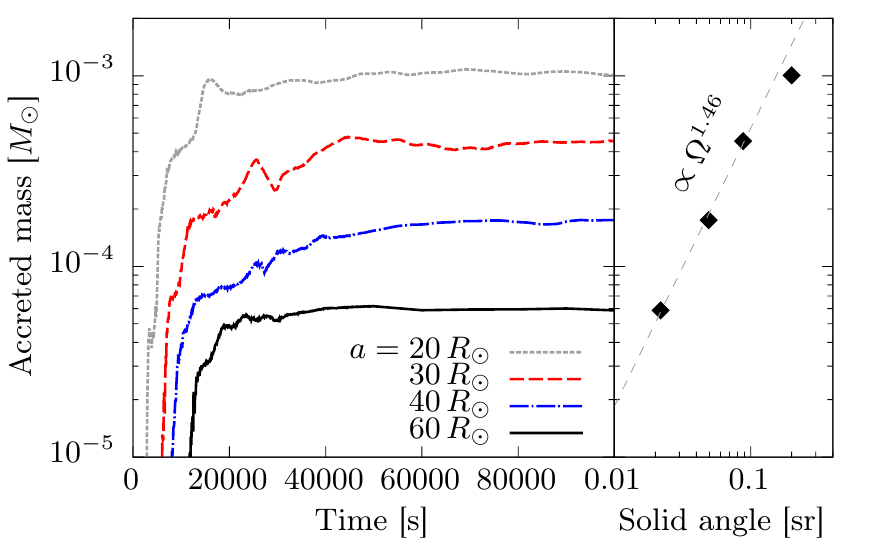}
 \caption{Time evolution of the accreted mass (left panel) and its dependence on solid angle (right panel). Dashed line in the right panel is a power law fit to the plots excluding the $a=20\,\rsun$ model.\label{fig:accretion}}
\end{figure}

Another possible way to probe the effects of ECI from observations is to see the surface pollution of heavy elements on the companion. In our simulations with tracer particles, some ejecta particles become bound to the star (red particles in Figure \ref{fig:tracer_snapshots}). By summing the masses of these bound ejecta particles, we can set a rough idea of the amount of pollution. In Figure \ref{fig:accretion} we display the time evolution and separation dependence of the amount of accreted matter. Note that the time axis extends further than in all the other figures we have shown so far. It takes a longer time for the accreted mass to reach a steady value than it does for the removal of mass or impact velocity because most of the accreted particles originate from the slower ejecta that arrive later after the bulk of the ejecta have flown past. In the right panel we plot the final accreted mass against the intersected solid angle. Similarly to the removed mass and impact velocity, the results roughly obey a power law except for the closest model. When we fit our results with a function $\Delta M_\mathrm{acc}/\msun=C\cdot\tilde{\Omega}^\mu$, the fitting parameters are $C=0.616$ and $\mu=1.46$. This translates to $\delta=-2.93$ if we fit it with $\Delta M_\mathrm{acc}/\msun=A\cdot(a/R_2)^\delta$ as in \citet{liu15b}. This power is considerably steeper than was obtained in \citet{liu15b} for 0.9 and $3.5\,\msun$ companions ($\delta=-0.884,\,-1.18$ respectively), which is consistent with their steepening trend with companion mass. However, extra care should be taken with this comparison because the explosion profiles used in the simulations are different.

 It can be seen in the left bottom panel of Figure \ref{fig:tracer_snapshots} that, when the star has a convective envelope, the accreted particles can be dragged into the interior of the star on a convective turnover timescale. In those cases the abundance of heavy elements will be smeared out over the convective layer and thus the surface abundance will be reduced. For radiative stars like the case we are interested in now, the heavy elements can linger on the surface a bit longer until they are mixed inside by thermohaline mixing. Therefore there may be a higher possibility of being able to observe the post-ECI pollution for radiative stars.

The accretion mainly occurs from the slower ejecta material that arrive later, which originated in the deeper layers of the progenitor with heavier elements. Our 1D explosion simulations do not include any information of the composition, and thus no nuclear reactions are taken into account. It is possible to trace back which layer in the progenitor the particles originated from, but our 1D approximation for the explosion does not allow us to take into account Rayleigh-Taylor instabilities which can mix up elements across composition boundaries \citep[e.g.][]{yam91,kif03,ham10,ono13,won15}. Therefore, estimating the composition of the accreted material is clearly beyond the scope of this paper. Also as we have discussed in Section \ref{sec:convergence}, the amount of removed mass is overestimated with our standard resolution and hence the amount of accretion is expected to be non-converged too. The qualitative behaviour of our results should still hold, but further work is required to quantitatively understand the contamination by ECI.

Most of the unbound material should be escaping the star with a velocity of the order of the local escape velocity. Although the mass is small, this bit of material can have a density large enough to create some dust at the time it cools down to dust condensation temperatures. Also the matter that is slowed down or has its path blocked by the presence of the companion can also contribute to this dust formation. It is difficult to estimate how much dust can be created, how the dust will be distributed, and what kind of dust is condensated, but it can partly contribute to the extinction towards the remaining companion from a few hundred days after the SN. However, there is also the possibility that a hot companion can dissociate some of the dust with its high energy radiation \citep[]{koc17}. For the smaller separation models, there is a large amount of unbound material to potentially create dust and the post-ECI expansion of the companion is large enough to lower the surface temperature and thus reduce the amount of high energy emission that dissociates the dust. Both effects are inversely proportional to the separation and therefore there should be a strong dependence on separation whether the post-ECI companion can self-enshroud itself with dust. This may provide an individual constraint on pre-SN binary parameters from observations of dust in SNe.

\section{Conclusion}\label{sec:conclusion}
We have conducted a comprehensive study of ejecta-companion interaction in massive binary systems via hydrodynamical simulations. In our first step we have carried out spherically symmetric hydrodynamical simulations of the explosion of the primary star, varying the ejecta mass and explosion energy. We then map the data from the first step to an axisymmetrical cylindrical grid and perform two-dimensional hydrodynamical simulations of the SN ejecta impacting the companion star. In this step, we explore a vast parameter space varying the mass, radius and structure of the companion and also the orbital separation. 

From our simulations we evaluate the removed mass and impact velocity of the companion, which both turned out to be very small. Our wide range of simulations allowed us to better understand the whole process of ECI, which does not seem to match the simple analytical model proposed in \citet{whe75}. We have introduced an alternative framework to estimate the impact velocity for arbitrary explosion profiles and companion star models. The key idea is that the intrinsic efficiency of momentum transfer from the ejecta to the companion is $\eta\sim0.5$, but can effectively be reduced because the star is compressed during the passage of the ejecta and reduces the cross section. The radius up to which the star is compressed is roughly determined by where the pressure in the star balances the ram pressure of the ejecta. One can therefore calculate the degree of compression given a companion star model and ejecta profile, and reduce $\eta$ by a square of that factor (Equation \ref{eq:eta_anal}). For most main sequence stars we suggest that the value $\eta\sim1/3$ will be reasonable.

The impact velocities obtained in our simulations were very small compared to the orbital velocity and average observed NS kick velocities. Therefore the impact velocity has very little influence on the survivability of the binary. The runaway velocity of the companion for disrupted binaries can be affected by the impact, but also only by a few percent. Surviving binaries can have slightly higher systemic velocities than when ECI is not taken into account.

We find that the total energy injected into the star by the SN ejecta is roughly $\sim8$--$10\,\%$ of the intersected kinetic energy. This efficiency can be modelled in the same way as the momentum transfer efficiency (Equation \ref{eq:energyinjection}). The injected energy is distributed mainly near the surface layers and declines inverse proportionally to the mass from the surface. We mimic this energy injection with the stellar evolution code \texttt{MESA} and follow the long term post-ECI evolution. The star becomes more than an order of magnitude more luminous after ECI and then retains its original state after about $\sim1000$\,years. In some cases the star can also expand up to radii that can engulf the primary NS and may trigger some interesting events. 

We also estimate the amount of surface contamination of heavy elements from the distribution of tracer particles in our hydrodynamical simulations. The total accreted mass is a factor of a few larger than the results obtained by \citet{liu15b} for $M_2=0.9,\,3.5\,\msun$ companions, when compared at the same orbital separations. We also find that the dependence on separation is considerably steeper than their results.

\acknowledgments
The authors thank Lorne Nelson for sharing computing facilities. Computations were partially carried out on facilities managed by Calcul Qu\'ebec and Compute Canada. This work has been supported by the Japan Society for the Promotion of Science (JSPS) KAKENHI Grant Numbers JP16H03986 and JP17K18792. The work has also been supported by a Humboldt Research Award to PhP at the University of Bonn. RH was supported by the JSPS overseas research fellowship No.\,29-514.

\software{MESA \citep[v10108;][]{MESA1,MESA2,MESA3,MESA4}}

\clearpage
\startlongtable
\begin{deluxetable*}{lccchcCcCcc}
 \tablecaption{Results of the hydrodynamical simulations with main sequence companions.\label{tab:results}}
 \tablehead{ \colhead{Model Name} & \colhead{Explosion} & \colhead{$M_2$} & \colhead{$R_2$} & \nocolhead{structure} & \colhead{$a$} & \colhead{$M_\mathrm{ub}$} & \colhead{$v_\mathrm{im}$} & \colhead{$M_\mathrm{ub,an}$\tablenotemark{a}} & \colhead{$v_\mathrm{im,an}$\tablenotemark{a}} &\colhead{$\eta$}\\
  \colhead{}&\colhead{}&\colhead{[$\msun$]}&\colhead{[$\rsun$]}&\nocolhead{}&\colhead{$[\rsun]$}&\colhead{$[\msun]$}&\colhead{[km\,s$^{-1}$]} &\colhead{$[\msun]$} &\colhead{[km\,s$^{-1}$]}  }
 \startdata
  M10R5MSa20-7c & 7.1c & 10 & 5 & MS & 20 & 3.915\times10^{-2} & 8.532 & 1.835\times10^{-1} & 56.45 & 0.13\\
  M10R5MSa30-7c & 7.1c & 10 & 5 & MS & 30 & 2.064\times10^{-2} & 7.108 & 9.203\times10^{-2} & 28.86 & 0.25\\
  M10R5MSa40-7c & 7.1c & 10 & 5 & MS & 40 & 1.260\times10^{-2} & 4.519 & 5.566\times10^{-2} & 17.84 & 0.29\\
  M10R5MSa60-7c & 7.1c & 10 & 5 & MS & 60 & 5.607\times10^{-3} & 2.200 & 2.694\times10^{-2} & 8.798 & 0.31\\
  M10R6MSa20-7c & 7.1c & 10 & 6 & MS & 20 & 6.233\times10^{-2} & 12.98 & 2.617\times10^{-1} & 71.74 & 0.14\\
  M10R6MSa30-7c & 7.1c & 10 & 6 & MS & 30 & 2.894\times10^{-2} & 9.963 & 1.339\times10^{-1} & 37.85 & 0.24\\
  M10R6MSa40-7c & 7.1c & 10 & 6 & MS & 40 & 1.645\times10^{-2} & 6.273 & 8.197\times10^{-2} & 23.51 & 0.28\\
  M10R6MSa60-7c & 7.1c & 10 & 6 & MS & 60 & 7.291\times10^{-3} & 2.958 & 4.025\times10^{-2} & 11.81 & 0.29\\
  M10R7MSa20-7c & 7.1c & 10 & 7 & MS & 20 & 9.821\times10^{-2} & 18.34 & 3.488\times10^{-1} & 86.97 & 0.14\\
  M10R7MSa30-7c & 7.1c & 10 & 7 & MS & 30 & 3.996\times10^{-2} & 12.95 & 1.817\times10^{-1} & 46.69 & 0.23\\
  M10R7MSa40-7c & 7.1c & 10 & 7 & MS & 40 & 2.060\times10^{-2} & 8.142 & 1.124\times10^{-1} & 29.60 & 0.26\\
  M10R7MSa60-7c & 7.1c & 10 & 7 & MS & 60 & 9.529\times10^{-3} & 3.819 & 5.595\times10^{-2} & 15.05 & 0.28\\
  M10R8MSa20-7c & 7.1c & 10 & 8 & MS & 20 & 1.554\times10^{-1} & 24.00 & 4.428\times10^{-1} & 102.6 & 0.14\\
  M10R8MSa30-7c & 7.1c & 10 & 8 & MS & 30 & 5.587\times10^{-2} & 15.98 & 2.343\times10^{-1} & 55.58 & 0.22\\
  M10R8MSa40-7c & 7.1c & 10 & 8 & MS & 40 & 2.706\times10^{-2} & 10.10 & 1.465\times10^{-1} & 35.72 & 0.25\\
  M10R8MSa60-7c & 7.1c & 10 & 8 & MS & 60 & 1.243\times10^{-2} & 4.751 & 7.378\times10^{-2} & 18.40 & 0.26\\
  M15R5MSa20-7c & 7.1c & 15 & 5 & MS & 20 & 3.480\times10^{-2} & 5.342 & 1.640\times10^{-1} & 42.65 & 0.13\\
  M15R5MSa30-7c & 7.1c & 15 & 5 & MS & 30 & 1.811\times10^{-2} & 5.076 & 8.008\times10^{-2} & 21.30 & 0.27\\
  M15R5MSa40-7c & 7.1c & 15 & 5 & MS & 40 & 1.099\times10^{-2} & 3.231 & 4.765\times10^{-2} & 12.84 & 0.31\\
  M15R5MSa60-7c & 7.1c & 15 & 5 & MS & 60 & 4.938\times10^{-3} & 1.522 & 2.262\times10^{-2} & 6.179 & 0.33\\
  M15R6MSa20-7c & 7.1c & 15 & 6 & MS & 20 & 5.275\times10^{-2} & 9.128 & 2.400\times10^{-1} & 55.85 & 0.15\\
  M15R6MSa30-7c & 7.1c & 15 & 6 & MS & 30 & 2.574\times10^{-2} & 7.305 & 1.192\times10^{-1} & 28.25 & 0.27\\
  M15R6MSa40-7c & 7.1c & 15 & 6 & MS & 40 & 1.508\times10^{-2} & 4.573 & 7.167\times10^{-2} & 17.20 & 0.30\\
  M15R6MSa60-7c & 7.1c & 15 & 6 & MS & 60 & 6.497\times10^{-3} & 2.139 & 3.446\times10^{-2} & 8.465 & 0.32\\
  M15R7MSa20-7c & 7.1c & 15 & 7 & MS & 20 & 8.129\times10^{-2} & 13.88 & 3.246\times10^{-1} & 68.18 & 0.16\\
  M15R7MSa30-7c & 7.1c & 15 & 7 & MS & 30 & 3.456\times10^{-2} & 9.678 & 1.639\times10^{-1} & 35.24 & 0.26\\
  M15R7MSa40-7c & 7.1c & 15 & 7 & MS & 40 & 1.827\times10^{-2} & 6.013 & 9.959\times10^{-2} & 21.84 & 0.29\\
  M15R7MSa60-7c & 7.1c & 15 & 7 & MS & 60 & 8.418\times10^{-3} & 2.774 & 4.847\times10^{-2} & 10.87 & 0.30\\
  M15R8MSa20-7c & 7.1c & 15 & 8 & MS & 20 & 1.231\times10^{-1} & 18.19 & 4.167\times10^{-1} & 80.80 & 0.16\\
  M15R8MSa30-7c & 7.1c & 15 & 8 & MS & 30 & 4.541\times10^{-2} & 12.17 & 2.136\times10^{-1} & 42.39 & 0.25\\
  M15R8MSa40-7c & 7.1c & 15 & 8 & MS & 40 & 2.281\times10^{-2} & 7.563 & 1.310\times10^{-1} & 26.57 & 0.28\\
  M15R8MSa60-7c & 7.1c & 15 & 8 & MS & 60 & 1.077\times10^{-2} & 3.489 & 6.449\times10^{-2} & 13.37 & 0.29\\
  M20R6MSa20-7c & 7.1c & 20 & 6 & MS & 20 & 4.826\times10^{-2} & 6.892 & 2.195\times10^{-1} & 45.56 & 0.15\\
  M20R6MSa30-7c & 7.1c & 20 & 6 & MS & 30 & 2.290\times10^{-2} & 5.751 & 1.069\times10^{-1} & 22.60 & 0.28\\
  M20R6MSa40-7c & 7.1c & 20 & 6 & MS & 40 & 1.344\times10^{-2} & 3.572 & 6.359\times10^{-2} & 13.62 & 0.31\\
  M20R6MSa60-7c & 7.1c & 20 & 6 & MS & 60 & 5.969\times10^{-3} & 1.655 & 3.016\times10^{-2} & 6.555 & 0.33\\
  M20R7MSa20-7c & 7.1c & 20 & 7 & MS & 20 & 7.059\times10^{-2} & 10.78 & 3.017\times10^{-1} & 56.55 & 0.17\\
  M20R7MSa30-7c & 7.1c & 20 & 7 & MS & 30 & 3.096\times10^{-2} & 7.727 & 1.492\times10^{-1} & 28.59 & 0.28\\
  M20R7MSa40-7c & 7.1c & 20 & 7 & MS & 40 & 1.710\times10^{-2} & 4.769 & 8.954\times10^{-2} & 17.40 & 0.31\\
  M20R7MSa60-7c & 7.1c & 20 & 7 & MS & 60 & 7.726\times10^{-3} & 2.184 & 4.295\times10^{-2} & 8.504 & 0.32\\
  M20R8MSa20-7c & 7.1c & 20 & 8 & MS & 20 & 1.049\times10^{-1} & 14.68 & 3.906\times10^{-1} & 67.61 & 0.17\\
  M20R8MSa30-7c & 7.1c & 20 & 8 & MS & 30 & 3.983\times10^{-2} & 9.863 & 1.959\times10^{-1} & 34.37 & 0.27\\
  M20R8MSa40-7c & 7.1c & 20 & 8 & MS & 40 & 2.063\times10^{-2} & 6.053 & 1.186\times10^{-1} & 21.14 & 0.30\\
  M20R8MSa60-7c & 7.1c & 20 & 8 & MS & 60 & 9.736\times10^{-3} & 2.757 & 5.751\times10^{-2} & 10.51 & 0.31\\
  M20R9MSa20-7c & 7.1c & 20 & 9 & MS & 20 & 1.489\times10^{-1} & 18.88 & 4.874\times10^{-1} & 78.29 & 0.17\\
  M20R9MSa30-7c & 7.1c & 20 & 9 & MS & 30 & 5.185\times10^{-2} & 12.09 & 2.476\times10^{-1} & 40.34 & 0.26\\
  M20R9MSa40-7c & 7.1c & 20 & 9 & MS & 40 & 2.548\times10^{-2} & 7.336 & 1.512\times10^{-1} & 25.08 & 0.28\\
  M20R9MSa60-7c & 7.1c & 20 & 9 & MS & 60 & 1.210\times10^{-2} & 3.380 & 7.407\times10^{-2} & 12.61 & 0.30\\
  M10R5MSa20-7h & 7.1h & 10 & 5 & MS & 20 & 1.113\times10^{-1} & 24.72 & 5.439\times10^{-1} & 185.6 & 0.19\\
  M10R5MSa30-7h & 7.1h & 10 & 5 & MS & 30 & 4.202\times10^{-2} & 15.39 & 2.838\times10^{-1} & 97.03 & 0.27\\
  M10R5MSa40-7h & 7.1h & 10 & 5 & MS & 40 & 2.392\times10^{-2} & 9.345 & 1.763\times10^{-1} & 61.33 & 0.29\\
  M10R5MSa60-7h & 7.1h & 10 & 5 & MS & 60 & 1.069\times10^{-2} & 4.518 & 8.832\times10^{-2} & 42.73 & 0.32\\
  M10R6MSa20-7h & 7.1h & 10 & 6 & MS & 20 & 2.073\times10^{-1} & 33.90 & 7.358\times10^{-1} & 228.1 & 0.18\\
  M10R6MSa30-7h & 7.1h & 10 & 6 & MS & 30 & 6.667\times10^{-2} & 20.29 & 3.932\times10^{-1} & 122.6 & 0.25\\
  M10R6MSa40-7h & 7.1h & 10 & 6 & MS & 40 & 3.383\times10^{-2} & 12.64 & 2.479\times10^{-1} & 78.00 & 0.28\\
  M10R6MSa60-7h & 7.1h & 10 & 6 & MS & 60 & 1.447\times10^{-2} & 6.119 & 1.266\times10^{-1} & 40.81 & 0.30\\
  M10R7MSa20-7h & 7.1h & 10 & 7 & MS & 20 & 3.472\times10^{-1} & 43.07 & 9.377\times10^{-1} & 267.4 & 0.16\\
  M10R7MSa30-7h & 7.1h & 10 & 7 & MS & 30 & 1.099\times10^{-1} & 25.33 & 5.118\times10^{-1} & 146.8 & 0.22\\
  M10R7MSa40-7h & 7.1h & 10 & 7 & MS & 40 & 5.005\times10^{-2} & 15.90 & 3.272\times10^{-1} & 94.88 & 0.25\\
  M10R7MSa60-7h & 7.1h & 10 & 7 & MS & 60 & 1.861\times10^{-2} & 7.830 & 1.699\times10^{-1} & 50.45 & 0.28\\
  M10R8MSa20-7h & 7.1h & 10 & 8 & MS & 20 & 5.325\times10^{-1} & 52.02 & 1.146\times10^{0}  & 305.4 & 0.15\\
  M10R8MSa30-7h & 7.1h & 10 & 8 & MS & 30 & 1.823\times10^{-1} & 30.69 & 6.373\times10^{-1} & 169.7 & 0.21\\
  M10R8MSa40-7h & 7.1h & 10 & 8 & MS & 40 & 7.675\times10^{-2} & 19.39 & 4.125\times10^{-1} & 111.5 & 0.24\\
  M10R8MSa60-7h & 7.1h & 10 & 8 & MS & 60 & 2.552\times10^{-2} & 9.661 & 2.176\times10^{-1} & 60.30 & 0.27\\
  M20R6MSa20-7h & 7.1h & 20 & 6 & MS & 20 & 1.242\times10^{-1} & 21.24 & 7.330\times10^{-1} & 161.7 & 0.23\\
  M20R6MSa30-7h & 7.1h & 20 & 6 & MS & 30 & 4.681\times10^{-2} & 12.66 & 3.682\times10^{-1} & 81.51 & 0.31\\
  M20R6MSa40-7h & 7.1h & 20 & 6 & MS & 40 & 2.583\times10^{-2} & 7.551 & 2.234\times10^{-1} & 49.97 & 0.33\\
  M20R6MSa60-7h & 7.1h & 20 & 6 & MS & 60 & 1.169\times10^{-2} & 3.572 & 1.089\times10^{-1} & 24.79 & 0.35\\
  M20R9MSa20-7h & 7.1h & 20 & 9 & MS & 20 & 5.003\times10^{-1} & 44.58 & 1.431\times10^{0}  & 255.1 & 0.20\\
  M20R9MSa30-7h & 7.1h & 20 & 9 & MS & 30 & 1.538\times10^{-1} & 24.01 & 7.554\times10^{-1} & 134.4 & 0.26\\
  M20R9MSa40-7h & 7.1h & 20 & 9 & MS & 40 & 6.476\times10^{-2} & 14.54 & 4.732\times10^{-1} & 84.79 & 0.28\\
  M20R9MSa60-7h & 7.1h & 20 & 9 & MS & 60 & 2.254\times10^{-2} & 6.992 & 2.402\times10^{-1} & 44.04 & 0.30\\
  M10R5MSa20-3c & 3.2c & 10 & 5 & MS & 20 & 1.687\times10^{-2} & 4.934 & 1.420\times10^{-1} & 45.18 & 0.19\\
  M10R5MSa30-3c & 3.2c & 10 & 5 & MS & 30 & 9.103\times10^{-3} & 3.594 & 7.066\times10^{-2} & 23.04 & 0.31\\
  M10R5MSa40-3c & 3.2c & 10 & 5 & MS & 40 & 5.208\times10^{-3} & 2.246 & 4.248\times10^{-2} & 14.03 & 0.34\\
  M10R5MSa60-3c & 3.2c & 10 & 5 & MS & 60 & 2.429\times10^{-3} & 1.103 & 2.044\times10^{-2} & 6.909 & 0.38\\
  M10R6MSa20-3c & 3.2c & 10 & 6 & MS & 20 & 2.896\times10^{-2} & 7.427 & 2.025\times10^{-1} & 57.88 & 0.19\\
  M10R6MSa30-3c & 3.2c & 10 & 6 & MS & 30 & 1.276\times10^{-2} & 4.894 & 1.026\times10^{-1} & 30.21 & 0.29\\
  M10R6MSa40-3c & 3.2c & 10 & 6 & MS & 40 & 6.702\times10^{-3} & 3.042 & 6.239\times10^{-2} & 18.60 & 0.32\\
  M10R6MSa60-3c & 3.2c & 10 & 6 & MS & 60 & 3.004\times10^{-3} & 1.481 & 3.041\times10^{-2} & 9.324 & 0.36\\
  M10R7MSa20-3c & 3.2c & 10 & 7 & MS & 20 & 4.788\times10^{-2} & 10.24 & 2.702\times10^{-1} & 70.86 & 0.19\\
  M10R7MSa30-3c & 3.2c & 10 & 7 & MS & 30 & 1.760\times10^{-2} & 6.351 & 1.392\times10^{-1} & 37.57 & 0.28\\
  M10R7MSa40-3c & 3.2c & 10 & 7 & MS & 40 & 8.583\times10^{-3} & 3.901 & 8.550\times10^{-1} & 23.57 & 0.30\\
  M10R7MSa60-3c & 3.2c & 10 & 7 & MS & 60 & 3.855\times10^{-3} & 1.888 & 4.217\times10^{-2} & 11.87 & 0.33\\
  M10R8MSa20-3c & 3.2c & 10 & 8 & MS & 20 & 7.995\times10^{-2} & 15.99 & 3.438\times10^{-1} & 83.74 & 0.23\\
  M10R8MSa30-3c & 3.2c & 10 & 8 & MS & 30 & 2.508\times10^{-2} & 8.034 & 1.797\times10^{-1} & 45.14 & 0.27\\
  M10R8MSa40-3c & 3.2c & 10 & 8 & MS & 40 & 1.124\times10^{-2} & 4.804 & 1.114\times10^{-1} & 28.45 & 0.29\\
  M10R8MSa60-3c & 3.2c & 10 & 8 & MS & 60 & 4.936\times10^{-3} & 2.320 & 5.557\times10^{-2} & 14.62 & 0.31\\
  M15R5MSa20-3c & 3.2c & 15 & 5 & MS & 20 & 1.522\times10^{-2} & 3.430 & 1.287\times10^{-1} & 33.80 & 0.19\\
  M15R5MSa30-3c & 3.2c & 15 & 5 & MS & 30 & 7.790\times10^{-3} & 2.627 & 6.247\times10^{-2} & 16.75 & 0.34\\
  M15R5MSa40-3c & 3.2c & 15 & 5 & MS & 40 & 4.335\times10^{-3} & 1.621 & 3.703\times10^{-2} & 10.09 & 0.37\\
  M15R5MSa60-3c & 3.2c & 15 & 5 & MS & 60 & 2.115\times10^{-3} & 0.783 & 1.750\times10^{-2} & 4.851 & 0.41\\
  M15R6MSa20-3c & 3.2c & 15 & 6 & MS & 20 & 2.350\times10^{-2} & 5.494 & 1.873\times10^{-1} & 44.49 & 0.21\\
  M15R6MSa30-3c & 3.2c & 15 & 6 & MS & 30 & 1.104\times10^{-2} & 3.668 & 9.235\times10^{-2} & 22.32 & 0.33\\
  M15R6MSa40-3c & 3.2c & 15 & 6 & MS & 40 & 6.046\times10^{-3} & 2.237 & 5.527\times10^{-2} & 13.57 & 0.36\\
  M15R6MSa60-3c & 3.2c & 15 & 6 & MS & 60 & 2.690\times10^{-3} & 1.077 & 2.642\times10^{-2} & 6.629 & 0.39\\
  M15R7MSa20-3c & 3.2c & 15 & 7 & MS & 20 & 3.815\times10^{-2} & 7.726 & 2.527\times10^{-1} & 54.71 & 0.22\\
  M15R7MSa30-3c & 3.2c & 15 & 7 & MS & 30 & 1.488\times10^{-2} & 4.753 & 1.266\times10^{-1} & 28.01 & 0.31\\
  M15R7MSa40-3c & 3.2c & 15 & 7 & MS & 40 & 7.567\times10^{-3} & 2.897 & 7.647\times10^{-2} & 17.21 & 0.34\\
  M15R7MSa60-3c & 3.2c & 15 & 7 & MS & 60 & 3.399\times10^{-3} & 1.385 & 3.696\times10^{-2} & 8.494 & 0.37\\
  M15R8MSa20-3c & 3.2c & 15 & 8 & MS & 20 & 5.856\times10^{-2} & 10.29 & 3.247\times10^{-1} & 65.42 & 0.22\\
  M15R8MSa30-3c & 3.2c & 15 & 8 & MS & 30 & 1.983\times10^{-2} & 5.971 & 1.649\times10^{-1} & 33.95 & 0.30\\
  M15R8MSa40-3c & 3.2c & 15 & 8 & MS & 40 & 9.382\times10^{-3} & 3.579 & 1.005\times10^{-1} & 21.08 & 0.32\\
  M15R8MSa60-3c & 3.2c & 15 & 8 & MS & 60 & 4.263\times10^{-3} & 1.715 & 4.909\times10^{-2} & 10.52 & 0.35\\
  M20R6MSa20-3c & 3.2c & 20 & 6 & MS & 20 & 2.116\times10^{-2} & 4.335 & 1.735\times10^{-1} & 36.18 & 0.23\\
  M20R6MSa30-3c & 3.2c & 20 & 6 & MS & 30 & 9.845\times10^{-3} & 2.903 & 8.407\times10^{-2} & 17.81 & 0.34\\
  M20R6MSa40-3c & 3.2c & 20 & 6 & MS & 40 & 5.393\times10^{-3} & 1.775 & 4.980\times10^{-2} & 10.66 & 0.38\\
  M20R6MSa60-3c & 3.2c & 20 & 6 & MS & 60 & 2.448\times10^{-3} & 0.850 & 2.352\times10^{-2} & 5.126 & 0.41\\
  M20R7MSa20-3c & 3.2c & 20 & 7 & MS & 20 & 3.283\times10^{-2} & 6.288 & 2.371\times10^{-1} & 45.16 & 0.24\\
  M20R7MSa30-3c & 3.2c & 20 & 7 & MS & 30 & 1.340\times10^{-2} & 3.831 & 1.165\times10^{-1} & 22.64 & 0.33\\
  M20R7MSa40-3c & 3.2c & 20 & 7 & MS & 40 & 6.975\times10^{-3} & 2.303 & 6.959\times10^{-2} & 13.67 & 0.36\\
  M20R7MSa60-3c & 3.2c & 20 & 7 & MS & 60 & 3.084\times10^{-3} & 1.103 & 3.320\times10^{-2} & 6.635 & 0.39\\
  M20R8MSa20-3c & 3.2c & 20 & 8 & MS & 20 & 4.922\times10^{-2} & 8.390 & 3.063\times10^{-1} & 53.98 & 0.24\\
  M20R8MSa30-3c & 3.2c & 20 & 8 & MS & 30 & 1.750\times10^{-2} & 4.783 & 1.525\times10^{-1} & 27.40 & 0.32\\
  M20R8MSa40-3c & 3.2c & 20 & 8 & MS & 40 & 8.368\times10^{-3} & 2.873 & 9.189\times10^{-2} & 16.71 & 0.34\\
  M20R8MSa60-3c & 3.2c & 20 & 8 & MS & 60 & 3.815\times10^{-3} & 1.369 & 4.427\times10^{-2} & 8.243 & 0.37\\
  M20R9MSa20-3c & 3.2c & 20 & 9 & MS & 20 & 7.231\times10^{-2} & 10.80 & 3.810\times10^{-1} & 62.47 & 0.24\\
  M20R9MSa30-3c & 3.2c & 20 & 9 & MS & 30 & 2.283\times10^{-2} & 5.864 & 1.921\times10^{-1} & 32.14 & 0.31\\
  M20R9MSa40-3c & 3.2c & 20 & 9 & MS & 40 & 1.068\times10^{-2} & 3.478 & 1.166\times10^{-1} & 19.95 & 0.33\\
  M20R9MSa60-3c & 3.2c & 20 & 9 & MS & 60 & 2.273\times10^{-2} & 6.997 & 5.671\times10^{-2} & 9.940 & 0.35\\
  M10R5MSa20-3h & 3.2h & 10 & 5 & MS & 20 & 9.519\times10^{-2} & 18.74 & 4.045\times10^{-1} & 146.5 & 0.23\\
  M10R5MSa30-3h & 3.2h & 10 & 5 & MS & 30 & 2.722\times10^{-2} & 10.66 & 2.087\times10^{-1} & 77.16 & 0.30\\
  M10R5MSa40-3h & 3.2h & 10 & 5 & MS & 40 & 1.337\times10^{-2} & 6.370 & 1.287\times10^{-1} & 48.29 & 0.32\\
  M10R5MSa60-3h & 3.2h & 10 & 5 & MS & 60 & 6.269\times10^{-3} & 3.085 & 6.378\times10^{-2} & 24.44 & 0.35\\
  M10R6MSa20-3h & 3.2h & 10 & 6 & MS & 20 & 2.008\times10^{-1} & 26.42 & 5.525\times10^{-1} & 182.3 & 0.22\\
  M10R6MSa30-3h & 3.2h & 10 & 6 & MS & 30 & 5.483\times10^{-2} & 14.32 & 2.915\times10^{-1} & 96.95 & 0.28\\
  M10R6MSa40-3h & 3.2h & 10 & 6 & MS & 40 & 2.221\times10^{-2} & 8.558 & 1.823\times10^{-1} & 62.05 & 0.30\\
  M10R6MSa60-3h & 3.2h & 10 & 6 & MS & 60 & 8.292\times10^{-3} & 4.133 & 9.195\times10^{-2} & 32.09 & 0.33\\
  M10R7MSa20-3h & 3.2h & 10 & 7 & MS & 20 & 3.464\times10^{-1} & 35.04 & 7.102\times10^{-1} & 214.5 & 0.21\\
  M10R7MSa30-3h & 3.2h & 10 & 7 & MS & 30 & 1.109\times10^{-1} & 18.43 & 3.824\times10^{-1} & 117.5 & 0.26\\
  M10R7MSa40-3h & 3.2h & 10 & 7 & MS & 40 & 4.015\times10^{-2} & 10.93 & 2.421\times10^{-1} & 75.68 & 0.28\\
  M10R7MSa60-3h & 3.2h & 10 & 7 & MS & 60 & 1.169\times10^{-2} & 5.308 & 1.241\times10^{-1} & 39.75 & 0.31\\
  M10R8MSa20-3h & 3.2h & 10 & 8 & MS & 20 & 5.471\times10^{-1} & 44.10 & 8.749\times10^{-1} & 248.6 & 0.20\\
  M10R8MSa30-3h & 3.2h & 10 & 8 & MS & 30 & 1.894\times10^{-1} & 23.13 & 4.795\times10^{-1} & 137.6 & 0.25\\
  M10R8MSa40-3h & 3.2h & 10 & 8 & MS & 40 & 7.273\times10^{-2} & 13.57 & 3.072\times10^{-1} & 89.19 & 0.26\\
  M10R8MSa60-3h & 3.2h & 10 & 8 & MS & 60 & 1.767\times10^{-2} & 6.507 & 1.597\times10^{-1} & 47.99 & 0.29\\
  M20R6MSa20-3h & 3.2h & 20 & 6 & MS & 20 & 9.915\times10^{-2} & 16.85 & 5.391\times10^{-1} & 126.6 & 0.29\\
  M20R6MSa30-3h & 3.2h & 20 & 6 & MS & 30 & 2.870\times10^{-2} & 8.931 & 2.686\times10^{-1} & 63.82 & 0.35\\
  M20R6MSa40-3h & 3.2h & 20 & 6 & MS & 40 & 1.451\times10^{-2} & 5.195 & 1.621\times10^{-1} & 38.82 & 0.36\\
  M20R6MSa60-3h & 3.2h & 20 & 6 & MS & 60 & 6.793\times10^{-3} & 2.439 & 7.843\times10^{-2} & 19.11 & 0.38\\
 \enddata
 \tablenotetext{a}{Analytical estimates of $M_\mathrm{ub}$ and $v_\mathrm{im}$ based on the model by \citet{whe75}.}
\end{deluxetable*}

\clearpage
\startlongtable
\begin{deluxetable*}{lcccccCchhc}
 \tablecaption{Results of the hydrodynamical simulations with polytrope companions.\label{tab:polytroperesults}}
 \tablehead{ \colhead{Model name} & \colhead{Explosion} & \colhead{$M_2$} & \colhead{$R_2$} & \colhead{Polytropic} & \colhead{$a$} & \colhead{$M_\mathrm{ub}$} & \colhead{$v_\mathrm{im}$} & \nocolhead{$M_\mathrm{ub,an}$} & \nocolhead{$v_\mathrm{im,an}$} & \colhead{$\eta$} \\
  \colhead{}&\colhead{}&\colhead{[$\msun$]}&\colhead{[$\rsun$]}&\colhead{Index}&\colhead{$[\rsun]$}&\colhead{$[\msun]$}&\colhead{[km\,s$^{-1}$]} &\nocolhead{$[\msun]$} &\nocolhead{[km\,s$^{-1}$]} & }
 \startdata
  M10R5P0a20-7c & 7.1c &10 & 5 & 0 & 20 &  5.090\times10^{-2} & 29.09 & 3.674\times10^{-1} & 136.8 & 0.45\\
  M10R5P0a30-7c & 7.1c &10 & 5 & 0 & 30 &  2.190\times10^{-2} & 15.93 & 1.633\times10^{-1} & 59.53 & 0.57\\
  M10R5P0a40-7c & 7.1c &10 & 5 & 0 & 40 &  1.262\times10^{-2} & 8.952 & 9.185\times10^{-2} & 33.24 & 0.57\\
  M10R5P0a60-7c & 7.1c &10 & 5 & 0 & 60 &  6.486\times10^{-3} & 3.884 & 4.082\times10^{-2} & 14.70 & 0.56\\
  M10R8P0a20-7c & 7.1c &10 & 8 & 0 & 20 &  1.779\times10^{0} & 109.5 & \times10^{-2} & & 0.54\\
  M10R8P0a30-7c & 7.1c &10 & 8 & 0 & 30 &  1.263\times10^{-1} & 45.46 & \times10^{-2} & & 0.62\\
  M10R8P0a40-3c & 3.2c &10 & 8 & 0 & 40 &  3.866\times10^{-2} &  24.33 & \times10^{-2} & & 0.60\\
  M10R8P0a60-3c & 3.2c &10 & 8 & 0 & 60 &  1.536\times10^{-2} & 9.996 & \times10^{-2} & & 0.56\\
  M10R5P15a20-7c & 7.1c &10 & 5 & 1.5 & 20 &  6.045\times10^{-2} & 20.35 & \times10^{-2} & & 0.32\\
  M10R5P15a30-7c & 7.1c &10 & 5 & 1.5 & 30 &  2.661\times10^{-2} & 13.23 & \times10^{-2} & & 0.47\\
  M10R5P15a40-7c & 7.1c &10 & 5 & 1.5 & 40 &  1.429\times10^{-2} & 7.696 & \times10^{-2} & & 0.49\\
  M10R5P15a60-7c & 7.1c &10 & 5 & 1.5 & 60 &  6.857\times10^{-3} & 3.292 & \times10^{-2} & & 0.47\\
  M10R8P15a20-7c & 7.1c &10 & 8 & 1.5 & 20 &  4.840\times10^{-1} & 76.01 & \times10^{-2} & & 0.43\\
  M10R8P15a30-7c & 7.1c &10 & 8 & 1.5 & 30 &  8.125\times10^{-2} & 35.91 & \times10^{-2} & & 0.49\\
  M10R8P15a40-7c & 7.1c &10 & 8 & 1.5 & 40 &  3.891\times10^{-2} & 19.90 & \times10^{-2} & & 0.49\\
  M10R8P15a60-7c & 7.1c &10 & 8 & 1.5 & 60 &  1.757\times10^{-2} & 8.597 & \times10^{-2} & & 0.48\\
  M10R5P3a20-7c & 7.1c &10 & 5 & 3 & 20 &  4.465\times10^{-2} & 10.69 & \times10^{-2} & & 0.17\\
  M10R5P3a30-7c & 7.1c &10 & 5 & 3 & 30 &  2.274\times10^{-2} & 8.306 & \times10^{-2} & & 0.29\\
  M10R5P3a40-7c & 7.1c &10 & 5 & 3 & 40 &  1.305\times10^{-2} & 5.180 & \times10^{-2} & & 0.33\\
  M10R5P3a60-7c & 7.1c &10 & 5 & 3 & 60 &  5.767\times10^{-3} & 2.386 & \times10^{-2} & & 0.34\\
  M10R8P3a20-7c & 7.1c &10 & 8 & 3 & 20 &  2.427\times10^{-1} & 35.23 & \times10^{-2} & & 0.20\\
  M10R8P3a30-7c & 7.1c &10 & 8 & 3 & 30 &  7.153\times10^{-2} & 21.39 & \times10^{-2} & & 0.29\\
  M10R8P3a40-7c & 7.1c &10 & 8 & 3 & 40 &  3.131\times10^{-2} & 12.98 & \times10^{-2} & & 0.32\\
  M10R8P3a60-7c & 7.1c &10 & 8 & 3 & 60 &  1.426\times10^{-2} & 5.900 & \times10^{-2} & & 0.33\\
  M10R5P0a20-3c & 3.2c &10 & 5 & 0 & 20 &  1.942\times10^{-2} & 14.25 & \times10^{-2} & & 0.54\\
  M10R5P0a30-3c & 3.2c &10 & 5 & 0 & 30 &  7.617\times10^{-3} & 7.007 & \times10^{-2} & & 0.60\\
  M10R5P0a40-3c & 3.2c &10 & 5 & 0 & 40 &  4.209\times10^{-3} & 3.928 & \times10^{-2} & & 0.60\\
  M10R5P0a60-3c & 3.2c &10 & 5 & 0 & 60 &  2.121\times10^{-3} & 1.783 & \times10^{-2} & & 0.62\\
  M10R8P0a20-3c & 3.2c &10 & 8 & 0 & 20 &  1.075\times10^{-1} & 50.56 & \times10^{-2} & & 0.72\\
  M10R8P0a30-3c & 3.2c &10 & 8 & 0 & 30 &  2.608\times10^{-2} & 19.69 & \times10^{-2} & & 0.65\\
  M10R8P0a40-3c & 3.2c &10 & 8 & 0 & 40 &  1.085\times10^{-2} & 10.40 & \times10^{-2} & & 0.62\\
  M10R8P0a60-3c & 3.2c &10 & 8 & 0 & 60 &  4.398\times10^{-3} & 4.385 & \times10^{-2} & & 0.59\\
  M10R5P15a20-3c & 3.2c &10 & 5 & 1.5 & 20 &  2.578\times10^{-2} & 11.19 & \times10^{-2} & & 0.42\\
  M10R5P15a30-3c & 3.2c &10 & 5 & 1.5 & 30 &  1.045\times10^{-2} & 6.072 & \times10^{-2} & & 0.52\\
  M10R5P15a40-3c & 3.2c &10 & 5 & 1.5 & 40 &  5.535\times10^{-3} & 3.477 & \times10^{-2} & & 0.53\\
  M10R5P15a60-3c & 3.2c &10 & 5 & 1.5 & 60 &  2.491\times10^{-3} & 1.575 & \times10^{-2} & & 0.55\\
  M10R8P15a20-3c & 3.2c &10 & 8 & 1.5 & 20 &  2.079\times10^{-1} & 39.37 & \times10^{-2} & & 0.56\\
  M10R8P15a30-3c & 3.2c &10 & 8 & 1.5 & 30 &  3.978\times10^{-2} & 16.44 & \times10^{-2} & & 0.54\\
  M10R8P15a40-3c & 3.2c &10 & 8 & 1.5 & 40 &  1.568\times10^{-2} & 8.990 & \times10^{-2} & & 0.54\\
  M10R8P15a60-3c & 3.2c &10 & 8 & 1.5 & 60 &  5.674\times10^{-3} & 3.945 & \times10^{-2} & & 0.53\\
  M10R5P3a20-3c & 3.2c &10 & 5 & 3 & 20 &  2.003\times10^{-2} & 6.073 & \times10^{-2} & & 0.23\\
  M10R5P3a30-3c & 3.2c &10 & 5 & 3 & 30 &  9.732\times10^{-3} & 4.090 & \times10^{-2} & & 0.35\\
  M10R5P3a40-3c & 3.2c &10 & 5 & 3 & 40 &  5.118\times10^{-3} & 2.497 & \times10^{-2} & & 0.38\\
  M10R5P3a60-3c & 3.2c &10 & 5 & 3 & 60 &  2.368\times10^{-3} & 1.177 & \times10^{-2} & & 0.41\\
  M10R5P3a400-3c & 3.2c & 10 & 5 & 3 & 400 & 6.669\times10^{-5} & 0.042 & \times10^{-2} & & 0.65\\
  M10R8P3a20-3c & 3.2c &10 & 8 & 3 & 20 &  1.130\times10^{-1} & 20.12 & \times10^{-2} & & 0.29\\
  M10R8P3a30-3c & 3.2c &10 & 8 & 3 & 30 &  2.915\times10^{-2} & 10.62 & \times10^{-2} & & 0.35\\
  M10R8P3a40-3c & 3.2c &10 & 8 & 3 & 40 &  1.328\times10^{-2} & 6.103 & \times10^{-2} & & 0.36\\
  M10R8P3a60-3c & 3.2c &10 & 8 & 3 & 60 &  5.582\times10^{-3} & 2.806 & \times10^{-2} & & 0.38\\
 \enddata
\end{deluxetable*}

\bibliographystyle{aasjournal}

\listofchanges
\end{document}